\documentclass[a4paper,fleqn,usenatbib]{mnras}

\usepackage[T1]{fontenc}
\usepackage{ae,aecompl}
\usepackage{rotating}

\usepackage{graphicx}	
\usepackage{amsmath}	
\usepackage{amssymb}	
\usepackage{float}
\usepackage[section]{placeins}
\usepackage{graphics}
\usepackage{verbatim}

\title[Exploiting Morphological Data from PWNe.]{Exploiting Morphological Data from Pulsar Wind Nebulae via a Spatio-Temporal Leptonic Transport Code}

\author[C. van Rensburg et al.]{
C. van Rensburg$^{1}$\thanks{E-mail: carlo.rensburg@gmail.com}, C. Venter$^{1}$, AS Seyffert$^{1}$ and Alice K. Harding$^2$
\\
$^{1}$Centre for Space Research, North-West University, Potchefstroom Campus, Private Bag X6001, Potchefstroom, South Africa, 2520\\
$^2$ Astrophysics Science Division, NASA Goddard Space Flight Center, Greenbelt, MD 20771, USA
}

\date{Accepted 2019 December 18. Received 2019 November 16; in original form 2019 August 19}

\pubyear{2019}

\begin{document}
\label{firstpage}
\pagerange{\pageref{firstpage}--\pageref{lastpage}}
\maketitle

\begin{abstract}
The next era of ground-based Cherenkov telescope development will see a great increase in both quantity and quality of $\gamma$-ray morphological data. This initiated the development of a spatio-temporal leptonic transport code to model pulsar wind nebulae. We present results from this code that predicts the evolution of the leptonic particle spectrum and radiation at different radii in a spherically-symmetric source. We simultaneously fit the overall broadband spectral energy distribution, the surface brightness profile and the X-ray photon index vs.\ radius for PWN 3C~58, PWN G21.5$-$0.9 and PWN G0.9+0.1. Such concurrent fitting of disparate data sets is non-trivial and we thus investigate the utility of different goodness-of-fit statistics, specifically the traditional $\chi^2$ test statistic and a newly developed scaled-flux-normalised test statistic to obtain best-fit parameters. We find reasonable fits to the spatial and spectral data of all three sources, but note some remaining degeneracies that will have to be broken by future observations.
\end{abstract}

\begin{keywords}
radiation mechanisms:non-thermal $-$ astroparticle physics $-$ methods: numerical $-$ ISM: supernova remnants $-$ gamma rays: general
\end{keywords}



\section{Introduction}
The 2020s will be one of the most exciting decades for very-high-energy (VHE, $E >$ 100 GeV) observational science as it will see the development of the new Cherenkov Telescope Array (CTA) with sites in the northern (19 telescopes) and the southern (100 telescopes) hemispheres \citep{CTA2019}. This array of telescopes, with its order-of-magnitude increase in sensitivity and near doubling in angular resolution compared to current telescopes such as H.E.S.S., VERITAS and MAGIC, will discover many more (older and fainter) $\gamma$-ray sources compared to the current population and reveal many more morphological details of currently known sources. One of these source classes is pulsar wind nebulae (PWNe). These sources are true multi-wavelength objects, observable from the highest $\gamma$-ray energies down to the radio waveband. Currently, there are 224 known VHE $\gamma$-ray sources, of which 35 are PWNe\footnote{http://tevcat2.uchicago.edu/}. The \textit{Fermi} Large Area Telescope (LAT) instrument has detected 5 high-energy $\gamma$-ray PWNe and 11 PWN candidates \citep{3FGL2015}. In the X-ray band there are 85 PWNe or PWN candidates, 71 of which have associated pulsars \citep{Kargaltsev2012}. 

The envisioned improvement in the CTA's angular resolution provides impetus for model development. The current modelling landscape can be summed up using three main categories, each with its own advantages and shortcomings. The first category encompasses magneto-hydrodynamic (MHD) codes \citep[e.g.,][]{Bucciantini2014, Porth2014, SlanePWN2017, Olmi2019} that are able to model the morphology (e.g., particle densities, bulk flows and magnetic profiles) of PWNe in great detail, but can not produce radiation spectra from first principles due to the fluid nature of these codes. Conversely, emission codes \citep[mostly leptonic; see, e.g.,][]{VdeJager2007,Zhang_2008,Tanaka_Takahara_2011,Martin2012,Martin2014,Torres2014}, are able to reproduce the radiation spectra reasonably well, but fail to model the PWN morphology as most of these codes model the source as a single sphere (0D) with parameters describing the PWN environment averaged over space. The third category encompasses hybrid models that combine either MHD or dynamical modelling results with those of an emission code to model both the morphology and radiation spectra. For example, \citet{Porth2016} use a steady-state transport code to predict the spatial emission properties of PWNe, but neglect the dynamical evolution of these sources. Also in this category are works where authors use semi-analytic expressions or numerical solutions to study the dynamical and radiation evolution of a PWNe inside its surrounding supernova remnants (see, e.g., \citealt{Gelfand2009} and references therein. While the emission codes mentioned above typically model young PWNe during their free-expansion phase, \citet{Martin2016,Torres2018,Torres2019} improve on this by considering the pressure produced by the particles and magnetic field inside the PWN, thus allowing them to model the interaction of the PWN and the supernova remnant (SNR) via the reverse shock during the reverberation phase of the PWN's evolution. Using such a model, \citet{Zhu2018} performed a population study including 18 PWNe, ranging from young to older PWNe, to discover relationships between model parameters. See \citet{Gelfand2017} for a recent review of this class of models.
Given the characteristics of the above model categories, we thus perceive a void in the current modelling landscape that has not been investigated substantially -- self-consistent modelling of both the spatial (in 1D, 2D or 3D) and temporal (from pulsar birth to present) aspects of the PWN spectrum. We developed such a spatio-temporal leptonic transport code \mbox{\citep{CvR2018MNRAS_G09}} and discuss its application to morphological PWN data in this paper. A similar type of model has been developed independently by \citet{Lu2017_model} and they fitted similar young PWNe as us in \citet{Lu2017_3C}; we will compare our results in what follows.

Beyond extending our PWN emission model to include a spatial dimension, an additional challenge to overcome when concurrently fitting spectral and spatial data is to handle disparate sets characterised by different numbers of degrees of freedom or dissimilar relative errors. A common way of finding an optimal model fit is to apply a $\chi^2$ test statistic for each of the data sets and minimise their sum. This is, however, not always the best solution for the situation at hand. Given the statistically heterogeneous nature of the data sets we encounter, we implement three different methods, i.e., by-eye, $\chi^2$ and a scaled-flux normalised test statistic (A.S.\ Seyffert et al., in preparation), in order to obtain model fits that reproduce both the spectral and spatial data.

In this paper we discuss our leptonic, spatially-dependent transport code (1D) that models the behaviour of the particles injected by the pulsar into the nebula and calculates the radiation spectra at different radii from the centre of the system (Section~\ref{sec:model}). We use three different methods of data fitting as discussed in Section~\ref{sec:method}. We fit three observables: the spectral energy distribution (SED), X-ray surface brightness (SB) and the X-ray photon index vs.\ radius for three PWN sources, i.e., PWN 3C~58, PWN G21.5$-$0.9 and PWN G0.9+0.1 (Section~\ref{sec:Results}). Our conclusions follow in Section~\ref{sec:concl}.

\section{The Model}\label{sec:model}

\subsection{The Code of \citet{CvR2018MNRAS_G09}}
Here we summarise the key highlights of our model. For full details, see \citet{CvR2018MNRAS_G09}. We solve the following transport equation:
\begin{equation}
\begin{split}
\frac{\partial N_{\rm{e}}}{\partial t} =& -\mathbf{V} \cdot (\nabla N_{\rm{e}}) +  \kappa \nabla^2 N_{\rm{e}} + \frac{1}{3}(\nabla \cdot \mathbf{V})\left( \left[\frac{\partial N_{\rm{e}}}{\partial \ln E_{\rm{e}}} \right] - 2N_{\rm{e}} \right)   \\
&+ \frac{\partial }{\partial E}(\dot{E}_{\rm{e,rad}}N_{\rm{e}}) +  Q(\mathbf{r},E_{\rm{e}},t),
\end{split}
\label{eq:transportFIN}
\end{equation} 
with $N_{\rm{e}}(\mathbf{r},E_{\rm{e}},t)$ the number of particles per unit energy and volume, \textbf{V} the bulk motion of particles, $\kappa$ the spatially-independent diffusion coefficient, $\dot{E}_{\rm{e,rad}}$ the total, i.e., synchrotron radiation (SR) and inverse Compton (IC) radiation energy loss rates, and $Q$ the particle injection spectrum with $r$ the radial dimension (assuming spherical symmetry) and $t$ the time since the PWN's birth. We  solve this transport equation including one spatial dimension. The particle injection spectrum is assumed to be a broken power law:
\begin{equation}
 Q(E_{\rm{e}},t) = \left\{\begin{matrix}
Q_0(t)\left(\frac{E_{\rm{e}}}{E_{\rm{b}}}\right)^{\alpha_1} \qquad E_{\rm{e,min}} \leq E_{\rm{e}}<E_{\rm{b}}\\ 
Q_0(t)\left(\frac{E_{\rm{e}}}{E_{\rm{b}}}\right)^{\alpha_2} \qquad E_{\rm{b}} < E_{\rm{e}} \leq E_{\rm{e,max}}
\end{matrix}\right.,
\label{brokenpowerlaw}
\end{equation}
with $Q_0(t)$ the time-dependent normalisation constant that is determined by equating the first moment of the injection spectrum to a constant fraction $\eta$ of the time-dependent pulsar spin-down luminosity, $E_{\rm{e}}$ the particle energy, $E_{\rm{b}}$ the break energy and $\alpha_1$ and $\alpha_2$ the spectral indices. To limit the number of free parameters in this model, we assume that $\alpha_1$ and $\alpha_2$ are time-independent. In the absence of an MHD code we chose to parametrise the bulk flow of the particles as well as the magnetic field profile. The bulk motion profile of the particles is parametrised according to $V(r) = V_0\left(r/r_0\right)^{\alpha_{\rm{V}}}$, with $V_0$ the bulk-flow normalisation, $r_0$ the inner (termination shock) radius of the PWN and $\alpha_{\rm{V}}$ the bulk-flow parameter. We parametrise the magnetic field as $B(r,t) = B_{\rm{age}}\left(r/r_0\right)^{\alpha_{\rm{B}}}\left(t/t_{\rm{age}}\right)^{\beta_{\rm{B}}},$ with $B_{\rm{age}}$ the present-day magnetic field at $r = r_0$ and $t = t_{\rm{age}}$ the PWN age. We assume Bohm-type diffusion: $\kappa(E_{\rm{e}}) = \kappa_B E_{\rm{e}}/B(r,t)$ and $\kappa_B = c/3e$, with $e$ denoting the elementary charge. We multiply $\kappa(E_{\rm{e}})$ by a scaling factor $\kappa_0$ to allow for diffusion that can be faster than what occurs in the very slow Bohm limit. 
With \citet{Kennel1984a} we assume that the magnetic field is toroidal and the bulk flow is purely radial. We also assume that, since the nebular plasma is a good conductor, we can apply ideal MHD equations (characterised by infinite macroscopic conductivity) to describe the PWN wind. In this case, Ohm's law becomes
\begin{equation}
    \mathbf{E} = -\frac{\mathbf{v}}{c}\times\mathbf{B}
\end{equation}
and by combining this with Faraday's law, we find \citep[e.g.,][]{Ferreira_deJager2008}
\begin{equation}
    \frac{\partial\mathbf{B}}{\partial t} = \nabla\times(\mathbf{v}\times\mathbf{B}).\label{eq:Faraday}
\end{equation}
To simply link the radial profiles of the magnetic field and bulk motion of the particles, we follow \citet{Schock2010,Holler12,Fang2017,Fang2019} and  assume that the temporal change in the magnetic field is slow enough\footnote{We have investigated this assumption \textit{a posteriori} for typical PWN parameters, notably $\beta_{\rm B}< -1$, and found that $\partial\mathbf{B}/\partial t < \nabla\times(\mathbf{v}\times\mathbf{B})$ for large values of $V_0$ and $\alpha_{\rm B}>0$ (implying larger spatial gradients) or large times $t$, small distances from the centre $r$, or small (more negative) $\beta_{\rm B}$ (implying smaller temporal gradients of the magnetic field) when  $\alpha_V + \alpha_B = -1$. This conclusion is independent of $B_{\rm age}$, but becomes stronger for smaller $r_0$ (larger spatial gradient) and larger $t_{\rm age}$ or larger (less negative) $\beta_{\rm B}$ (smaller temporal gradient). For small values of $\alpha_B$, similar to what we will be using in what follows, the spatial term dominates (to various degrees) the temporal one for most times and distances, especially later times and closer distances. Thus, we think that it is a reasonable approximation to drop the temporal term in favour of the spatial one for our current modelling. In the general case of Equation~(\ref{eq:Faraday}), one can show that $\alpha_V + \alpha_B + 1 \propto f(r,t)$, with $f$ representing some multivariate function (and $f=0$ if $\partial\mathbf{B}/\partial t=0$), i.e., in this general case one would not be able to use our parametric prescription for $\mathbf{V}$ or $\mathbf{B}$, as the indices do not remain constant with $t$ nor $r$. For a more general description (and including the magnetic field's temporal derivative in this linking equation), MHD modelling would be needed.} that we can set $\partial \mathbf{B}/\partial t \simeq 0$ in the above equation, even for a time-dependent prescription of $B$ (this assumption holds exactly true for steady-state models such as those of \citealt{Kennel1984a,Vorster2013}). From this follows $VBr=V_0 B_0 r_0={\rm constant}$, which for our parametric specifications of the magnetic field and bulk flow implies that 
\begin{equation}\label{eq:alpValpB}
 \alpha_V + \alpha_B = -1.
\end{equation}
This relation is used to reduce the number of free parameters by one and to simplify our search for best-fit parameters in later sections. 

We solve Equation \eqref{eq:transportFIN} numerically\footnote{In \citet{vRensburg2014} we demonstrated convergence of the model output as one increases the mesh density, i.e., showing convergence of both the particle spectrum and SED with an increase in the number of radial, particle energy and photon energy bins.
} and predict the following PWN properties: (1) the \textit{SED}; (2) radiation spectra are calculated for each zone, which in turn are used to perform a line-of-sight (LOS) calculation that yields the predicted \textit{SB profile}; (3) the SR component of the emitted SED for each zone is used to calculate the \textit{X-ray photon index vs.\ radial distance} from the embedded pulsar by fitting a power-law curve to the model SED in the 2.0$-$10.0~keV energy band. All three of these properties are fitted simultaneously and the best-fit parameters are given in the respective tables below.

\subsection{Energy Conservation}\label{sec:EC}
The rotational energy of the pulsar is ultimately the reservoir from which is derived the energy of the pulsar wind particles and electromagnetic fields \citep[e.g.,][]{Gelfand2017}. We explicitly direct a fraction $\eta$ of the spin-down luminosity to the particle injection spectrum as described above. 
We tested the consistency of our calculations by integrating the injection spectrum multiplied by particle energy $E_{\rm e}Q(E_{\rm e})$ over volume and particle energy  \citep[e.g.,][]{Sefako2003}, and did indeed recover $\eta$. 

Secondly, we have to ensure that the fraction of spin-down power being converted into magnetic energy amounts to no more than $\eta_{\rm B} = 1 - \eta\ll 1$.
\citet{Torres2014} and others explicitly ensure energy conservation by solving for the magnetic field using the following equation 
\begin{equation}
    \frac{dW_{\rm B}}{dt} = \eta_{\rm B} L - \frac{W_{\rm B}}{R_{\rm{PWN}}}\left( \frac{dR_{\rm{PWN}}}{dt} \right),
    \label{eq:dwdt}
\end{equation}
with the final term describing adiabatic losses, $W_{\rm B}$ being the magnetic energy \begin{equation}
    W_{\rm B}\equiv \frac{4\pi}{3}R_{\rm{PWN}}^3(t)\frac{B^2(t)}{8\pi},
    \label{eq:W_B}
\end{equation}
and assuming a PWN radius \citep{Swaluw2001b}
\begin{equation}
R_{\rm PWN}(t) = C\left(\frac{L_0 t}{E_0}\right)^{1/5}V_{\rm ej} t\propto t^{6/5},
\label{eq:R_PWN}
\end{equation}
with $C\approx 1$, $V_{\rm ej}$ the ejecta speed, $L_0$ the pulsar spin-down luminosity at birth, and  $E_0$ the energy of the supernova explosion.

\begin{figure}
  \centering
  \includegraphics[width=\columnwidth]{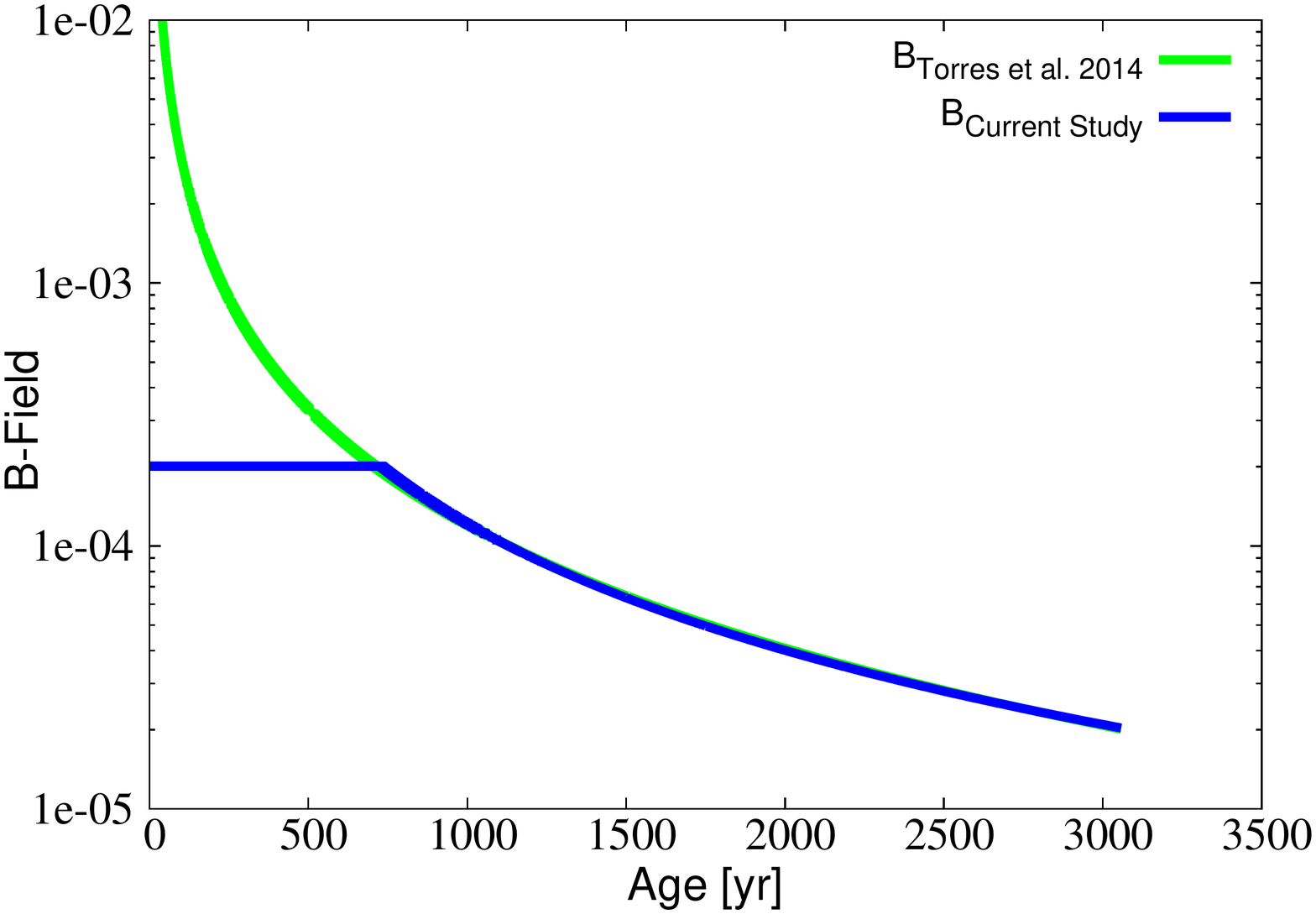}
  \caption{Comparison between the parametric form (blue line, $\alpha_{\rm B}=0$, $\beta_B=-1.6$) and the analytical form used by \citet{Torres2014} (green line) of the magnetic field of  PWN G0.9$+$0.1. Our magnetic field is set to a constant at early times to make the code more efficient (i.e., to limit the dynamical time step that scales as the SR loss time scale $t_{\rm SR}\propto B^{-2}$).\label{Fig:app_bt}}
\end{figure}
We used a parametric form for the magnetic field to allow us to explore a wider variety of magnetic field behaviour, given the uncertainty in PWN radius and $\eta_{\rm B}$ (considering, e.g., its time-dependence or not).
However, we implicitly ensured energy conservation by letting our magnetic field approximate the one calculated as explained above (see Figure~\ref{Fig:app_bt}). Since our respective magnetic fields are similar, so too are our predicted SR components (see Figure~3 of \citealt{CvR2018MNRAS_G09}).

\begin{figure}
  \centering
  \includegraphics[width=\columnwidth]{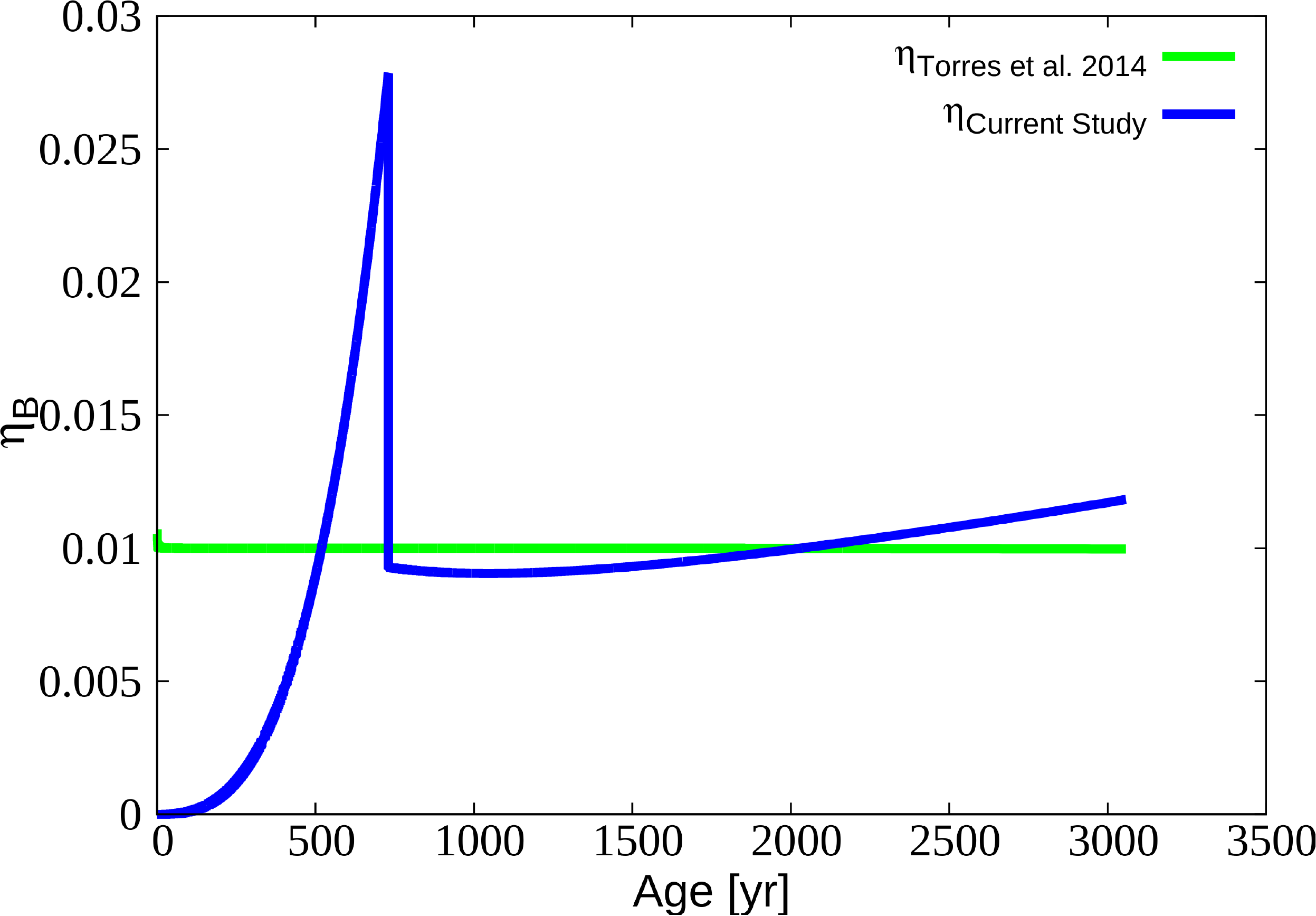}
  \caption{Comparison of calculated $\eta_{\rm B}$ for the parametric form (blue line) vs.\ the analytical form of the magnetic field used by \citet{Torres2014} (green line).\label{Fig:app_eta}}
\end{figure}

In order to address this point more quantitatively, we solved for $\eta_{\rm B}$ using Equation~(\ref{eq:dwdt}) upon substituting our parametric expression for the magnetic field (see Figure~\ref{Fig:app_eta}). We used Equation~(\ref{eq:R_PWN}) to approximate the PWN radius\footnote{In the expression for $W_{\rm B}$ given in Equation~(\ref{eq:W_B}), it was not clear what value to use for $R_{\rm PWN}$. This is because in our model, we assume a distant escaping boundary and let the particles flow into a static radial grid and radiate, so we may determine the time-dependent observed PWN size without having to impose it \textit{a priori}. Any radius derived from the observed size of the PWN is necessarily dependent on photon energy; also, it is arbitrary to define a radius based on some fixed drop in the steady-state particle spectrum. Thus, we opted to use the standard analytic expression for free expansion, which allowed us to directly compare our results to those of \citet{Torres2014}}. From Figure~\ref{Fig:app_eta}, we recover the correct value of $\eta_{\rm B}\approx1\%$ for the case of G0.9+0.1 \citet{Torres2014}. Our parametric magnetic field slightly underestimates the analytic magnetic field at early times, leading to a lower value of $\eta_{\rm B}$ at those times. At an age of approximately 800~years, the parametric magnetic field follows the $\beta_B=-1.6$ profile and thus the $\eta_{\rm B}$ value is similar to that of \citet{Torres2014}. The discontinuous behaviour around 800 years is a result of taking the derivative of a discontinuous magnetic field around this age. We thus found that our implied value for $\eta_{\rm B}$ is close to that of \citet{Torres2014}, demonstrating energy conservation in our model (since $\eta_{\rm B}\approx 0.01$ and $\eta\approx0.99$ for this source, so the particle energetics substantially dominates).

\subsection{Comparison to Other Spatial Models}
Similar to our work, \citet{Lu2017_model} models young PWNe ($<10$~kyr) that have not been influenced by the returning reverse shock. Their code solves a radially-dependent transport equation similar to ours, but coupled to a photon conservation equation, and additionally includes the effects of particle injection due to photon-photon pair production as well as synchrotron-self-absorption that decreases the number of escaping photons. They also consider synchrotron-self-Compton (SSC) emission. They find that photon-photon pair production should be negligible, and SSC is usually not that important compared to IC (except possibly for the Crab Nebula, see \citealt{Torres2014}).
They model the injection of particles in a similar fashion to our approach, using a toroidal magnetic field topology, and project the predicted radiation using a similar line-of-sight calculation.

There are also a few differences in our modelling approaches. \citet{Lu2017_model} invoke the definition of the outer boundary of the PWN as derived by \citet{Bucciantini2004} using their spherically-symmetric relativistic MHD PWN model. We, on the other hand, define an escaping boundary radius far beyond the expected PWN radius and allow the particles to freely expand from their initial injection site at the termination shock and escape at this distant boundary. Also, as noted before and discussed in Section~\ref{sec:EC}, we use a parametric form for the magnetic field and the bulk motion of the particles to be able to mimic MHD results and test a wider variety of possibilities, while they solve an equation to constrain the evolution of the magnetic field with time due to a changing pulsar spin-down luminosity and adiabatic expansion losses. We tested this method in \citet{CvR2018MNRAS_G09} and found that our results were not impacted greatly by the difference in approach. Finally, \citet{Lu2017_model} use a time-dependent diffusion coefficient, while our coefficient is independent of time. Lastly, they fix the bulk flow normalisation $V_0(\sigma)$ while we leave this as a free parameter. 

\citet{Porth2016} noted that it is important to extend the results of \citet{Tang2012} who modelled the profile of the X-ray photon index by including also the surface-brightness profile. \citet{Porth2016} thus studied the spatial X-ray properties of G21.9$-$0.9, 3C~58 and the inner parts of the Vela PWN using three different codes: a three-dimensional MHD code, test-particle simulations as well as a Fokker-Planck particle transport model. The MHD flow provided the background geometry in which they could determine the diffusion coefficients using the test-particle code. They focused on diffusion in a turbulent flow due to eddies on the scale of the termination shock radius, finding that such a diffusion coefficient is quite independent of particle energy.
Their transport model is similar to the one used in this paper, but is a steady-state model invoking spherically symmetric transport including convection, diffusion, adiabatic cooling, and SR losses. It takes as input time-averaged Gaussian-fit radial profiles for the magnetic field and bulk speed that are obtained from their 3D MHD model. It is not clear that these profiles are similar to ours, given our assumption for frozen-in magnetic flux. Also, they assume a time-independent diffusion coefficient which is dependent on radius but independent of energy, contrary to our assumption of energy dependence. \citet{Porth2016} lastly considered only the X-ray band, while \citet{Lu2017_3C} also considered the radio and TeV energy bands.

\section{Methods for Finding Best fits}\label{sec:method}
As mentioned in Section~\ref{sec:model}, the data sets (broadband SED, SB profile and X-ray photon index vs.\ radius) we use to constrain our spatial model are statistically heterogeneous, which causes some sets to dominate\footnote{The $\chi^2$ values of data sets with various different errors may vary substantially resulting in some data sets to be preferred by optimisations schemes.} others when inferring best-fit parameters. This may be due to the subsets having different numbers of data points (compare, e.g, the number of points in the SED vs.\ that of the spectral index profile), or to a discrepancy in respective relative errors, resulting in $\chi^2$ values that differ substantially between subsets. Thus, we decided to investigate different fitting methods to find best fits to the respective data sets associated with each PWN we considered. In this Section, three different fitting methods are discussed. We hope to break degeneracies in our model by considering different effective weightings of each data subset, while also testing a new type of statistic not previously used for PWN codes.

The first and most basic method is using trial and error and knowledge (intuition) of the code's behaviour upon changing parameters, using by-eye fitting\footnote{During this procedure, the parameters are changed more or less at the same time. A test was done to find best fits following different orders of changing the free parameters. However, this did not have any effect on the final best-fit parameters, since one has to change the parameters nearly simultaneously to find any sort of best fit, even when starting with different initial parameters. We thus conclude that our by-eye fitting procedure is robust, even if it is subjective.} to find a best model fit. This is a viable method, but is subjective and limited to the available time and resolution of searching, given a large parameter space. The uniqueness of the resulting best fit, and whether it is truly optimal, are also unknown. This is, however, a good starting point and for all the sources we modelled, we indicate these fits using black lines below.

The next way of finding a best fit is to perform a Pearson's $\chi^2$ test \citep{Bevington1969}. By calculating the test statistic
\begin{equation}
 \chi^2 = \sum_i\left(\frac{D_i-M_i}{\epsilon_i}\right)^2,
 \label{eq:x2}
\end{equation}
with $D_i$ the data point, $\epsilon_i$ the error and  $M_i$ the model value corresponding to the $i$th bin, and minimising this value over all free model parameters, one can find a best fit by combining the $\chi^2$ values for all the subsets (usually biased by one particular data subset), and minimising this composite quantity. Examples of this methodology are afforded by \citet{Porth2016}, who used a Nelder-Mead minimisation method coupled to a $\chi^2$ test statistic to find optimal parameters for their transport code when applying their model to three PWNe (fitting SB profiles and X-ray photon spectral index profiles). Similarly, \citet{Lu2017_model} applied their model to PWN 3C~58, PWN G21.5$-$0.9 and PWN MSH~15$-$52, using a $\chi^2$ method to first fit the SED and then predict the spatial behaviour of each source based on the best-fit parameters found using only the SED. Since we are not only fitting a single data set, we calculate the $\chi^2$ value for each of the subsets and then minimise the sum to find the best possible fit (as shown by the red lines in the figures below). This method seems to work reasonably well.

We find, however, that the heterogeneous nature of the data subsets implies that some subsets will dominate when using a standard $\chi^2$ test statistic. To ameliorate this, we tested a new best-fit method (A.S.\ Seyffert et al., in preparation) originally developed for dual-band pulsar light curve modelling, where the relative radio errors are usually much smaller than that of the $\gamma$-ray band \citep{Bezuidenhout2018}. The scaled-flux-normalised (SFN) test statistic $\chi^2_\Phi$ is a modified $\chi^2$ statistic that is better suited to handle multiple, statistically heterogeneous, binned data sets. The $\chi^2_\Phi$ statistic attempts to eliminate the dominance of certain data subsets by effectively rescaling the goodness-of-fit measure of each subset such that it reflects how well the model reproduces large-scale trends in the data, rather than how much the model deviates from the data points in terms of the data errors (as is done by the traditional $\chi^2$ method). This is done by considering a particular data set to be a perturbation above some background level\footnote{Choosing background levels is a bit problematic, but we used the following values during this initial application of the new method: 0.01 of the SED level, $\Gamma=3.0$ for the X-ray index background value and a constant value of 0.01 for the SB background level. Fortunately, different choices in background level did not significantly affect our best-fit values.} $\boldsymbol{B}$ and calculating the scaled flux $\Phi^2 = \chi^2 \left( \boldsymbol{B} \right)$ for each data subset. The latter is essentially the normal $\chi^2$ value for the background level. The SFN test statistic is defined as 
\begin{equation}
    \chi^2_{\Phi} \left( \boldsymbol{M} \right) \equiv \frac{\Phi^2 - \chi^2 \left( \boldsymbol{M} \right)}{\Phi^2 - \nu},
    \label{eq:sfn}
\end{equation}
where $\boldsymbol{M}$ is a given model prediction and $\nu$ is the number of degrees of freedom associated with the fit in that data subset's domain. For a model with $n_{\rm p}$ parameters being fit to a data set with $n$ bins, $\nu = n - n_{\rm p}$. Eq.~(\ref{eq:sfn}) will typically yield a value between $0$ and $1$, with 0 meaning the fit is as good as if the background were our model (i.e., no source present) and 1 meaning that it is within 1$\sigma$ from the data. Values outside this range are also possible, with $\chi^2_\Phi > 1$ indicating that the errors were over-estimated (i.e., overfitting) and $\chi^2_\Phi < 0$ indicating that the model fit was worse than a flat model at the background level. These two values are equivalent in meaning to having $\chi^2 < \nu$ and $\chi^2 > \Phi^2$ (the latter usually corresponding to $\chi^2/\nu\gg1$). When fitting 3 data subsets concurrently, we calculate the average of the three SFN values and maximise this to find the best possible fit. Our results using this method are indicated by the blue lines in Figures~\ref{Fig:3C58_SED} to \ref{Fig:G09_index}. 

Another idea that we tested to find optimal concurrent fits was to introduce a free nuisance parameter to optimise the amplitude of the SB as in Figures~\ref{Fig:3C58_SB}, \ref{Fig:G21_SB} and \ref{Fig:G09_SB}, since only the overall SB profile shape is modelled and not the absolute SB values. However, this did not yield significantly improved fits and therefore we did not implement this in our final fitting method. Furthermore, we previously showed that our code is able to predict the size of the PWN as a function of energy \citep{CvR2018POS}, but since the sizes are directly derived from the SB profiles for different energy bands, this size vs.\ photon energy output is not an independent observable. We thus decided to only fit the SB profiles and not additionally the size vs.\ energy, since the first are more fundamental and the second should directly derive from them.

Using the methods and types of data sets described above, we found suitable fits to the observed data of three PWNe, as discussed in the next Section. In Table~\ref{tbl:3C58} to~\ref{tbl:G09} the last three rows indicate that the optimal model parameters have been found using either the $\chi^2$ or $\chi^2_{\Phi}$ test statistic. The corresponding values of the same test statistic (per row), using best-fit parameters preferred by the other search methods, are shown for comparison.

\section{Results and discussion}\label{sec:Results}
We use PWN 3C~58, PWN G21.5$-$0.9 and PWN G0.9+0.1 as case studies. This choice of sources is due to the availability of radial data in the X-ray regime as well as existing modelling efforts by independent authors with which to compare our results with \citep[e.g.,][]{Slane2004, Matheson2005, 2012XMMG09}. For a more direct comparison we note that \citet{Lu2017_3C} predict the SED and X$-$ray photon index vs.\ radius for PWN 3C 58 and PWN G21.5$-$0.9, but not the SB profile; they did, however, predict all three of these observables for PWN MSH 15$-$52 \citep{Lu2017_model}. We found reasonable fits to the available spatial and spectral data. Figures \ref{Fig:3C58_SED} to \ref{Fig:3C58_index} show the results for PWN 3C 58, while Figures \ref{Fig:G21_SED} to \ref{Fig:G21_index} do the same for PWN G21.5$-$0.9 and Figures \ref{Fig:G09_SED} to \ref{Fig:G09_index} for PWN G0.9+0.1. 

In order to make our investigation tractable\footnote{We only specify the best-fit values of the free parameters and not the errors or error contours on these parameters. This is because a full Markov-Chain Monte Carlo (MCMC) approach (based on maximising the likelihood) requires substantial, even prohibitive computational resources. However, we did perform an MCMC procedure for a very crude (but still reasonable) spatial and energy resolution to study this issue in some more detail, and considered two cases. For the first case, we invoked very large ranges in the flat priors for our free parameters, finding several local maxima and very broad contours, but with one global maximum being preferred. This illustrates that one generally cannot simply quote linear (asymmetric) errors on free parameters in the tables below, since the contours are disconnected and encompass several maxima. In the second case, we zoomed into the parameter space by bracketing the priors around the best-fit parameters, finding that while the local maxima contours seem reasonable, in some cases physical limits have to be imposed such as a non-zero diffusion coefficient or bulk speed that does not exceed the speed of light. This illustrates the difficulty in estimating reasonable parameter errors, but also point to the fact that ideal (physically acceptable) fits that can concurrently reproduce all available data may not exist within the current model's parameter space.} in terms of computational facilities and available time, we freed the following five model parameters: the current-day magnetic field strength ($B_{\rm{age}}$), the bulk flow normalisation of particles ($V_0$), the age of the system ($t_{\rm{age}}$), the magnitude of the diffusion coefficient ($\kappa_0$) and the radial profile of the magnetic field ($\alpha_{\rm B}$). These parameters have the most significant influence on the radial properties of the PWN. As noted in Section~\ref{sec:model} the magnetic field's radial profile and the bulk motion of the particles' radial profile are coupled ($\alpha_B+\alpha_V = -1$), thus the bulk-flow normalisation is an independent parameter, but the bulk-flow profile is dependent on the choice of magnetic field profile. Below, we discuss each of the modelled sources individually as each of them posed their own unique challenges. 

\subsection{PWN 3C~58}
\cite{Weiler1971} discovered PWN 3C~58 and originally classified it as an SNR, but later radio observations by \cite{Weiler1978} showed a bright, centre-filled morphology as well as a flat radio spectrum leading to its re-classification as a PWN. Several decades later an associated pulsar, PSR J0205+6449, was discovered by \cite{Murray2002}, having a rotational period of $65$ ms and a spin-down luminosity of $2.7 \times 10^{37}$erg s$^{-1}$. Subsequent X-ray observations \citep{Slane2004} have shown filaments and knots that closely resemble those seen in the Crab Nebula and therefore PWN 3C 58 is characterised as being ``Crab-like''. \cite{Roberts1993} derived a distance of 3.2 kpc which we will use throughout this paper, while  \citet{Tanaka2013} suggest a closer distance of 2.0 kpc. The X-ray SB and radial photon index vs.\ radius were taken from \cite{Slane2004} as shown in Figures~\ref{Fig:3C58_SB} and \ref{Fig:3C58_index}. \citet{Abdo2013} observed PWN 3C 58 with \textit{Fermi-LAT} and detected a spectrum extending past 100 GeV, having a power-law spectral index of $\Gamma = 1.61 \pm 0.21$ with a flux of $(1.75 \pm 0.68) \times 10^{-11} \rm{erg}~\rm{cm}^{-2}\rm{s}^{-1}$. VHE data for PWN 3C 58 were obtained from \citet{Aleksic2014}. During their observations with MAGIC in the energy range between 400 GeV to 10 TeV they observed a flux of ($2.0\pm 0.4_{\rm{stat}}\pm 0.6_{\rm{sys}})\times 10^{-13}\rm{cm}^{-1}\rm{s}^{-1}\rm{TeV}^{-1}$ which is one of the lowest PWN flux measured to date. The spectrum is well described by a power-law function with $\Gamma = 2.4\pm 0.2_{\rm{stat}} \pm 0.2_{\rm{sys}}$. For more information regarding 3C~58 also see \cite{Li2018}.

\begin{figure}
  \centering
  \includegraphics[width=\columnwidth]{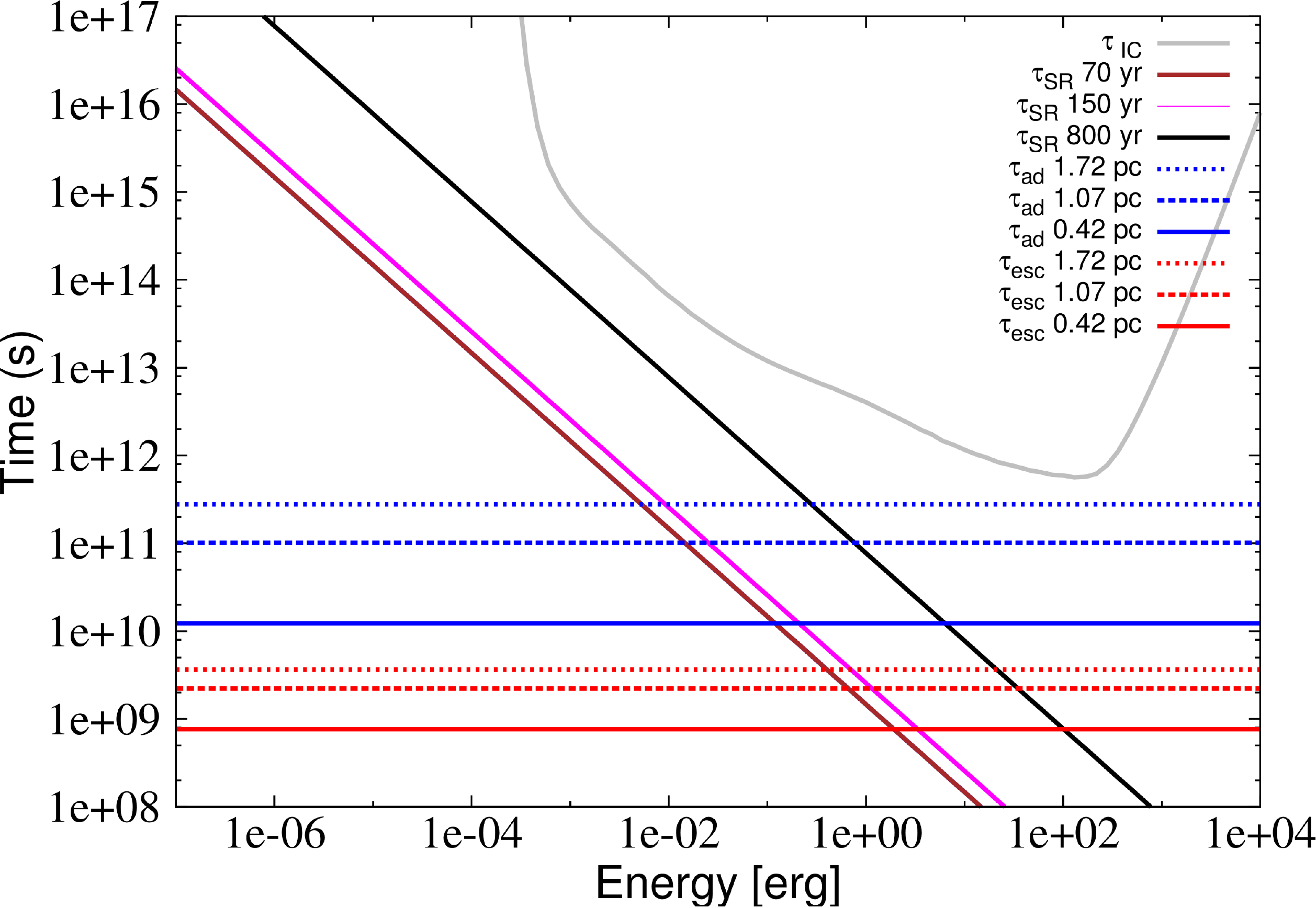}
    \caption{Timescales of the key processes in the model for PWN 3C~58 for different snapshots in time and at different radii. Here $\tau_{\rm{IC}}$ is the IC energy loss time, $\tau_{\rm{SR}}$ the SR loss time and is given for three different ages of the PWN, $\tau_{\rm{ad}}$ is the adiabatic loss time given for three different positions in the PWN and $\tau_{\rm{esc}}$ is the time for a particle to escape the current zone (spherical shell) in the PWN via bulk flow, also given for three different positions. Here $\tau_{\rm{IC}}$ is the IC energy loss time scale (independent of $t$ and $r$).\label{Fig:Timescales}}
\end{figure}

Figure~\ref{Fig:Timescales} shows the timescales of different processes in the PWN model. This is only shown for PWN 3C~58 to assist in the discussion, but is similar for the different modelled sources. Here $\alpha_{\rm B}=0$, implying a constant magnetic field vs.\ radius. This is an assumption made throughout the paper and is discussed in the next paragraph. From Figure~\ref{Fig:Timescales} we can clearly see that the escape timescale ($\tau_{\rm{esc}}$) is the shortest for all radii and all time for lower particle energies ($E_{\rm e}<1$ erg), indicating that such particles will generally leave the current zone in the model before losing a substantial amount of energy due to radiation losses. For energies $E_{\rm e}>1$ erg the SR energy-loss timescale ($\tau_{\rm{SR}}$) starts to dominate. These two effects influence the radial predictions from the code most significantly. The associated model quantities are the magnetic field (directly influencing $\tau_{\rm{SR}}$) and the normalised bulk flow of particles (determining the time spent in each zone by the particles, thus $\tau_{\rm{esc}}$).  

\begin{figure}
  \centering
  \includegraphics[width=\columnwidth]{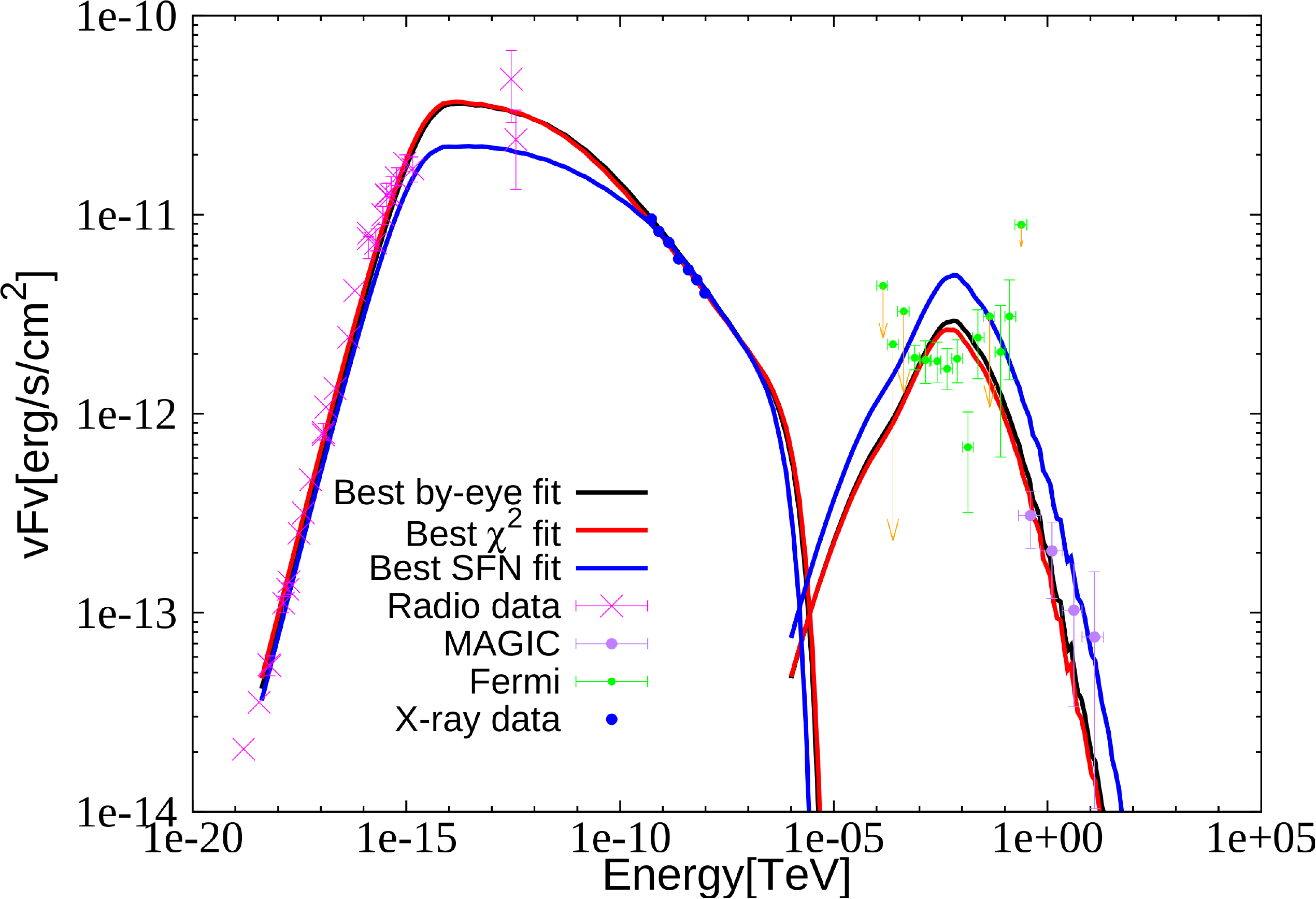}
    \caption{Broadband SED for PWN 3C 58, with radio data from \textit{WMAP} \citep{Weiland2011}, infrared data from \textit{IRAS} \citep[][]{Green1994,Slane2008}, X-ray data from \textit{ASCA} \citep{Torii2000}, GeV data from \textit{Fermi}-LAT \citep{Abdo2013} and TeV data from MAGIC \citep{Aleksic2014}.\label{Fig:3C58_SED}}
\end{figure}

\begin{figure}
  \centering
  \includegraphics[width=\columnwidth]{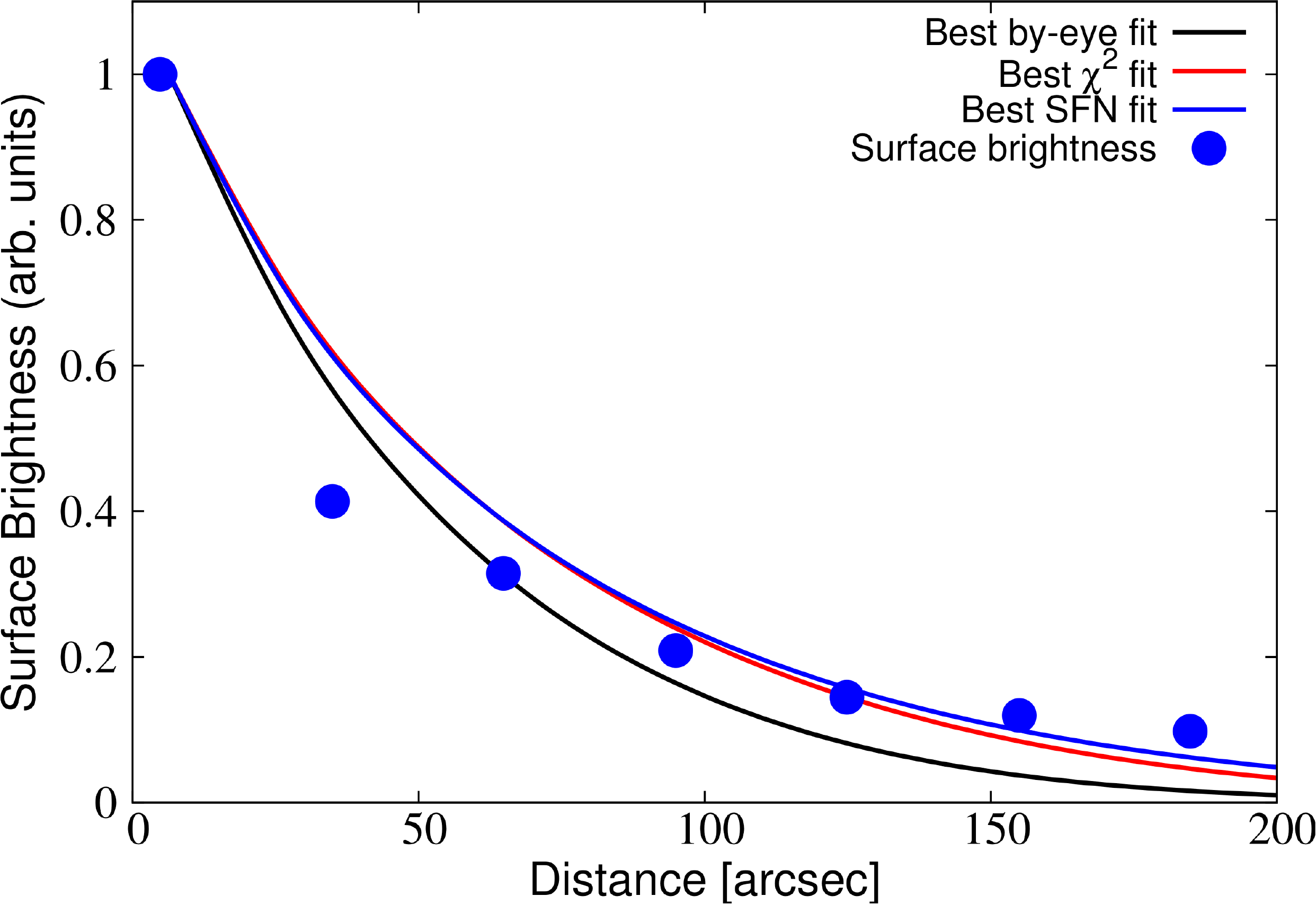}
     \caption{SB profile for PWN 3C 58. The data points are from \citet{Slane2004} for the energy range $0.5-4$~keV and the lines indicate our model best fits, with fitting methods indicated in the legend. \label{Fig:3C58_SB}}
\end{figure}

\begin{figure}
  \centering
  \includegraphics[width=\columnwidth]{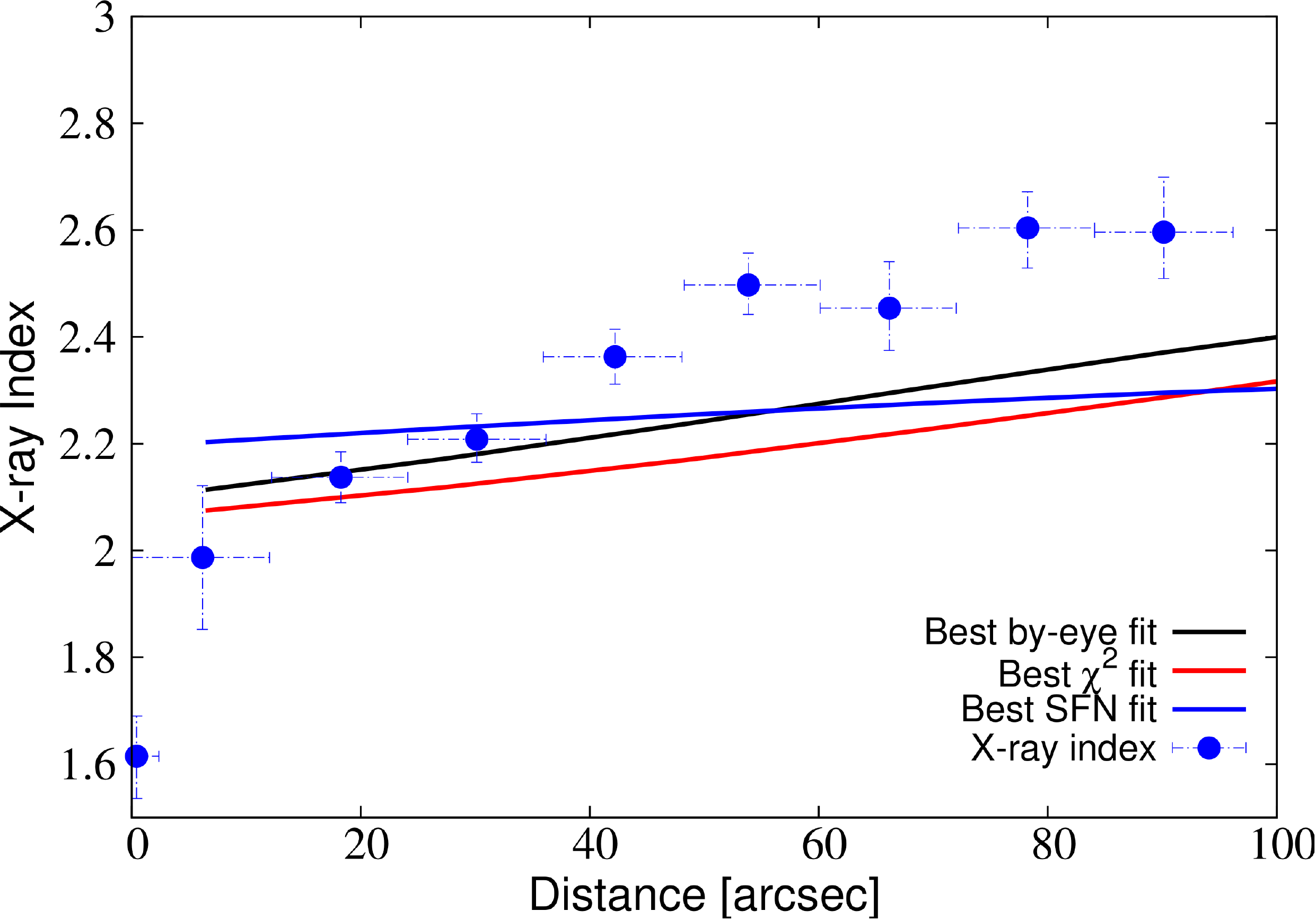}
  \caption{X-ray photon index for PWN 3C~58 vs.\ radius. The data points are from \citet{Slane2004} and the lines indicate our model best fits. \label{Fig:3C58_index}}
\end{figure}

From Figures~\ref{Fig:3C58_SED} to~\ref{Fig:3C58_index}, we note that for 3C~58 we find reasonable fits to all three the data subsets simultaneously and that all three fitting methods give similar results. Our best-fit parameters are similar to those found by \citet{Torres2013ApJ}.
Table~\ref{tbl:3C58} shows that the different methods gave similar best-fit parameter values, with the SFN method giving a smaller present-day magnetic field value, thus requiring an older age for the system to still be able to fit the SED. The age range found by all fitting methods is much lower than the pulsar characteristic age of $\tau_{\rm c}=5~380$~yr, perhaps pointing to a braking index $n>3$ and a birth period close to the current period ($P_0\approx P$). Interestingly, these preferred ages are much closer to the one argued for by \citet{Kothes2013} of 830~yr that is linked to the historical SN explosion in 1 181 A.D. The radial magnetic field profile parameter $\alpha_{\rm B}$ was initially chosen to be free, but all of the optimisation schemes yielded $\alpha_{\rm B} \sim 0$ and we therefore set $\alpha_{\rm B}=0$. This implies a constant magnetic field vs.\ radius and a velocity profile that goes like $V(r)\sim1/r$. This finding justifies the assumption made in 0D models that the magnetic field is constant throughout the PWN, and to some extent explains their success in modelling the SEDs of several sources \citep{Gelfand2009}. 
The fits indicate a large bulk flow of particles as well as a normalisation ($\kappa_0$) for the diffusion coefficient that exceeds the Bohm diffusion coefficient by two orders of magnitude. These values are, however, still viable except for a bulk velocity of $3.5\times10^{10}$ cm s$^{-1}$ found as the best fit by the $\chi^2$ method, which exceeds the speed of light. The combined $\chi^2$ values for all three methods are large, indicating a formally bad fit or an underestimation of errors (in the absence of published errors for the SB profile we assume a $10\%$ relative error). On the other hand, the $\chi^2_\Phi$ values are close to $1.0$ with the maximal value of $0.94$ characterising the best fit. The $\chi^2_\Phi$ values being so close to $1.0$ reflects the fact that this method uses units of scaled flux rather than errors, and that it balances the contributions of the different subsets.

Generally, for this source and G21.5$-$0.9, \citet{Porth2016} find similar values for the magnetic field, but smaller values for the bulk flow and higher values for the spatial diffusion coefficient. Our inferred values also seem similar to those of \citet{Lu2017_3C}. We attribute differences to the different model implementations.

\begin{table}
\resizebox{\columnwidth}{!}{%
\begin{tabular}{lrrrr}
\hline
\textit{Fixed parameters}					& 	 		& 		&		&   \\ \hline 
Period ($P$) (s)					& & & 0.065     		    		 		&   \\ 
Time derivative of period ($\dot P$) (s s$^{-1}$) 	& & & 1.5$\times10^{-13}$      		 		&   \\ 
Spin-down luminosity ($L_{\rm{age}}$) (erg/s)   	& & & 2.7$\times10^{37}$	  		 		&   \\ 
Braking index ($n$)               	        	& & & 3          		  		 		&   \\ 
Distance to the source (kpc)          			& & & 3.2          		  		 		&   \\ 
Index of the injected spectrum ($\alpha_1$) 		& & & 1.31         		  		 		&   \\ 
Index of the injected spectrum ($\alpha_2$) 		& & & 2.92          	  		 		&   \\ 
Break energy ($\gamma_{\rm{b}}$) 			& & & 9$\times10^{4}$         		 		&   \\ 
Conversion efficiency ($\eta$)               		& & & 0.2          		   		 		&   \\ 
Magnetic field time dependence ($\beta_{\rm B}$)		& & & $-1.0$      		   		 		&   \\
Soft-photon components:	& & $T$ (K) & $u$ (eV/cm$^3$) &       		   		 		   \\ 
\quad Cosmic microwave background (CMB)		& & 2.76 & 0.23       		   		 		&   \\
\quad Infrared  		& & 30.0 & 2.5      		   		 		&   \\
\quad Optical & & 3000.0 &  25.0     		   		 		&   \\ \hline\hline
\textit{Fitted parameters}				& 	 		 				&   \\ \hline 
							& By-eye 		& $\chi^2$ 	&	$\chi^2_\Phi$	&   \\ \hline 

Radial parameter of the magnetic field  ($\alpha_{\rm B}$)      	& 0.0          		& 0.0		&0.0		&   \\ 
Present-day magnetic field ($\mu$G)			& 65.7        		& 70.7		&39.1		&   \\ 
Bulk flow normalisation ($10^{10}$ cm s$^{-1}$)	& 1.8          		& 3.5		&1.0		&   \\ 
Age (kyr)                			& 1.121      		& 1.152	&1.589	&   \\ 
Diffusion coefficient normalisation ($\kappa_{\rm{0}}$) 				& 80.9         		& 133.1	&74.0		&   \\
$\chi^2/\rm{\nu}$ 			& 697/46 & 380/46	& 830/46		&   \\
 & (15.2) & (8.3)	& (18.0) &   \\
$\chi^2_\Phi$			& 0.88         		& 0.93		& 0.94			&   \\ \hline
\end{tabular}
}
\caption{Best-fit parameters for PWN 3C 58, with $T$ the temperature and $u$ the energy density assumed for each soft-photon blackbody component.\label{tbl:3C58}}
\end{table}

\subsection{PWN G21.5$-$0.9}
PSR J1833$-$1034 is one of the youngest pulsars in the Galaxy with an estimated age of 870 yr \citep{Bietenholz2008}. This pulsar powers PWN G21.5$-$0.9 that has a nearly spherical shape in the radio and X-ray bands. \cite{Camilo2006} found the pulsar period of 61.8 ms with a period derivative of 2.02~$\times$~10$^{-13}$~s~s$^{-1}$. The distance to PWN G21.5$-$0.9 is estimated to be 4.7$\pm$0.4~kpc \citep{Camilo2006,Tian2008}. VHE $\gamma$-rays in the 1$-$10 TeV energy range have been detected by the H.E.S.S.\ experiment during their Galactic Plane Survey and they found a flux of $2.6\times 10^{33}(d/4.9\rm{kpc})^2 \rm{erg~s}^{-1}$ and a power-law spectral index of $\Gamma= 2.42\pm 0.19$ \citep{HESSgps2018}.

\begin{figure}
  \centering
  \includegraphics[width=\columnwidth]{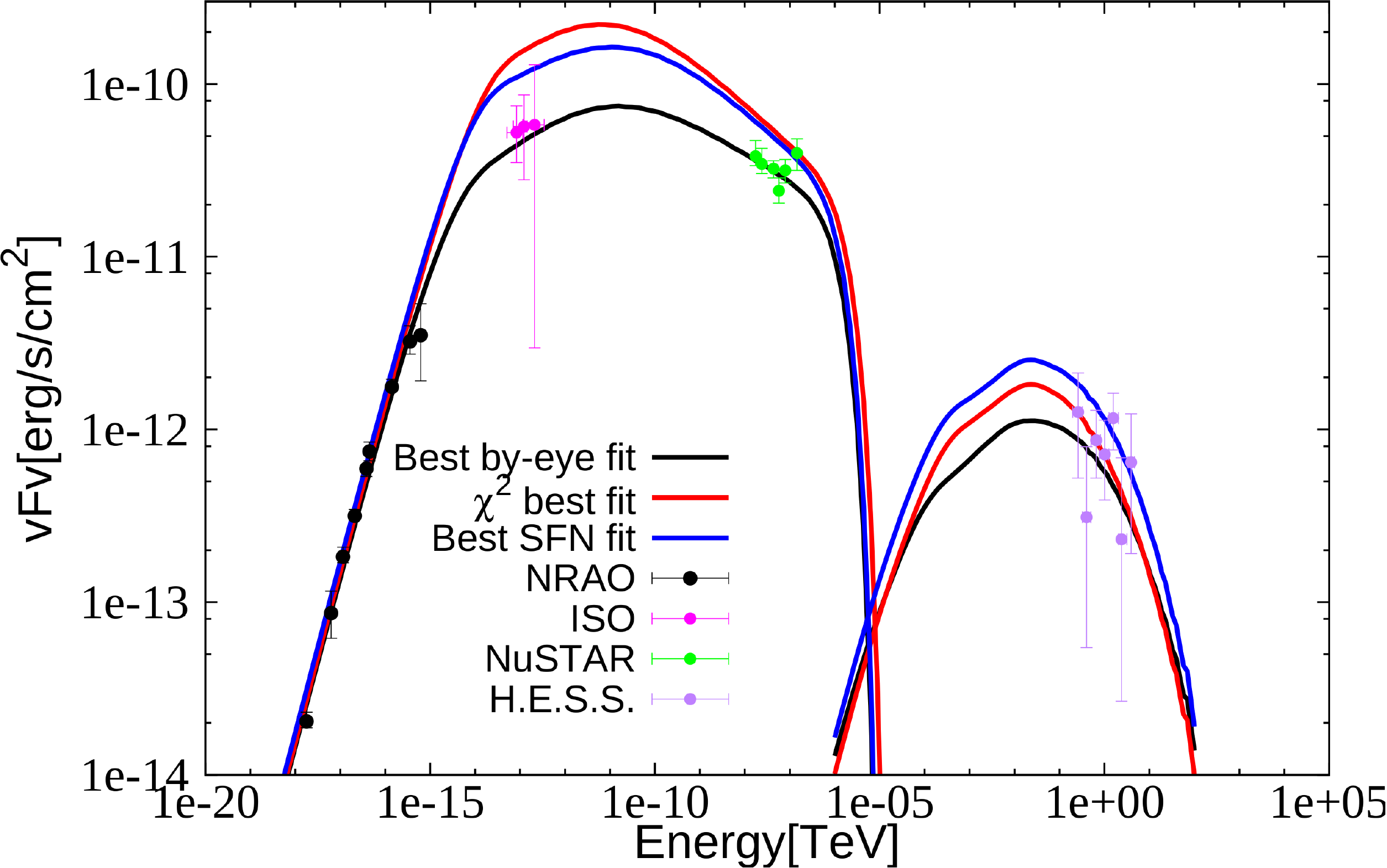}
  \caption{Broadband SED for PWN G21.5$-$0.9. The radio data are from NRAO observations \citep{Salter1989}, infrared data from the \textit{Infrared Space Observatory} \citep{Gallant1998}, X-rays from \textit{NuSTAR} observations by \citet{Nynka2014} and the TeV data from H.E.S.S. \citep{Djannati2008}.
  \label{Fig:G21_SED}}
\end{figure}

\begin{figure}
  \centering
  \includegraphics[width=\columnwidth]{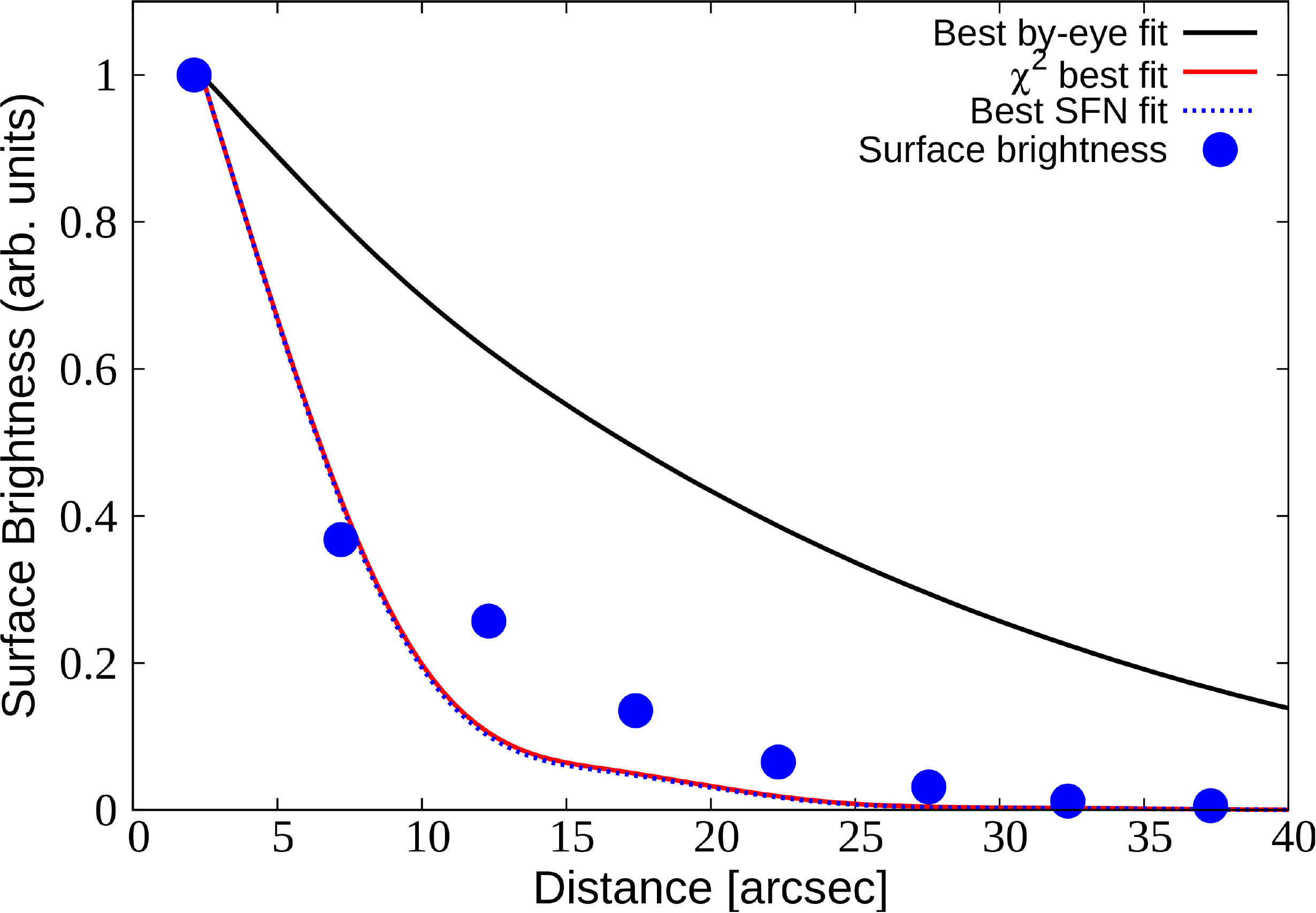}
  \caption{SB profile for PWN G21.5$-$0.9 with data points from \citet{Matheson2005} and the lines indicating the model best fits.
    \label{Fig:G21_SB}}
\end{figure}

\begin{figure}
  \centering
  \includegraphics[width=\columnwidth]{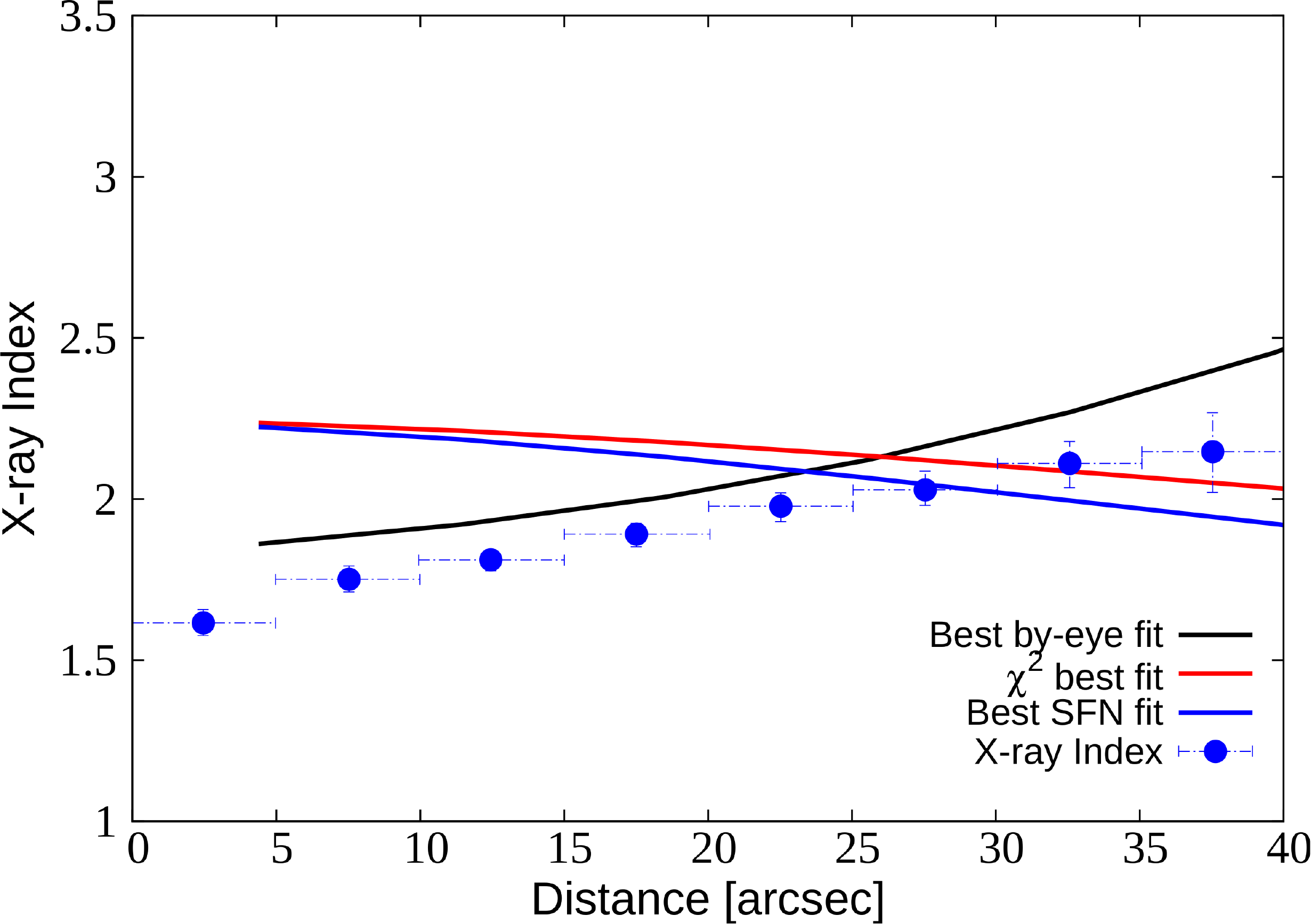}
  \caption{X-ray index profile for PWN G21.5$-$0.9 with data points from \citet{Matheson2005} and the lines indicating the model best fits.
    \label{Fig:G21_index}}
\end{figure}

\begin{figure}
  \centering
  \includegraphics[width=\columnwidth]{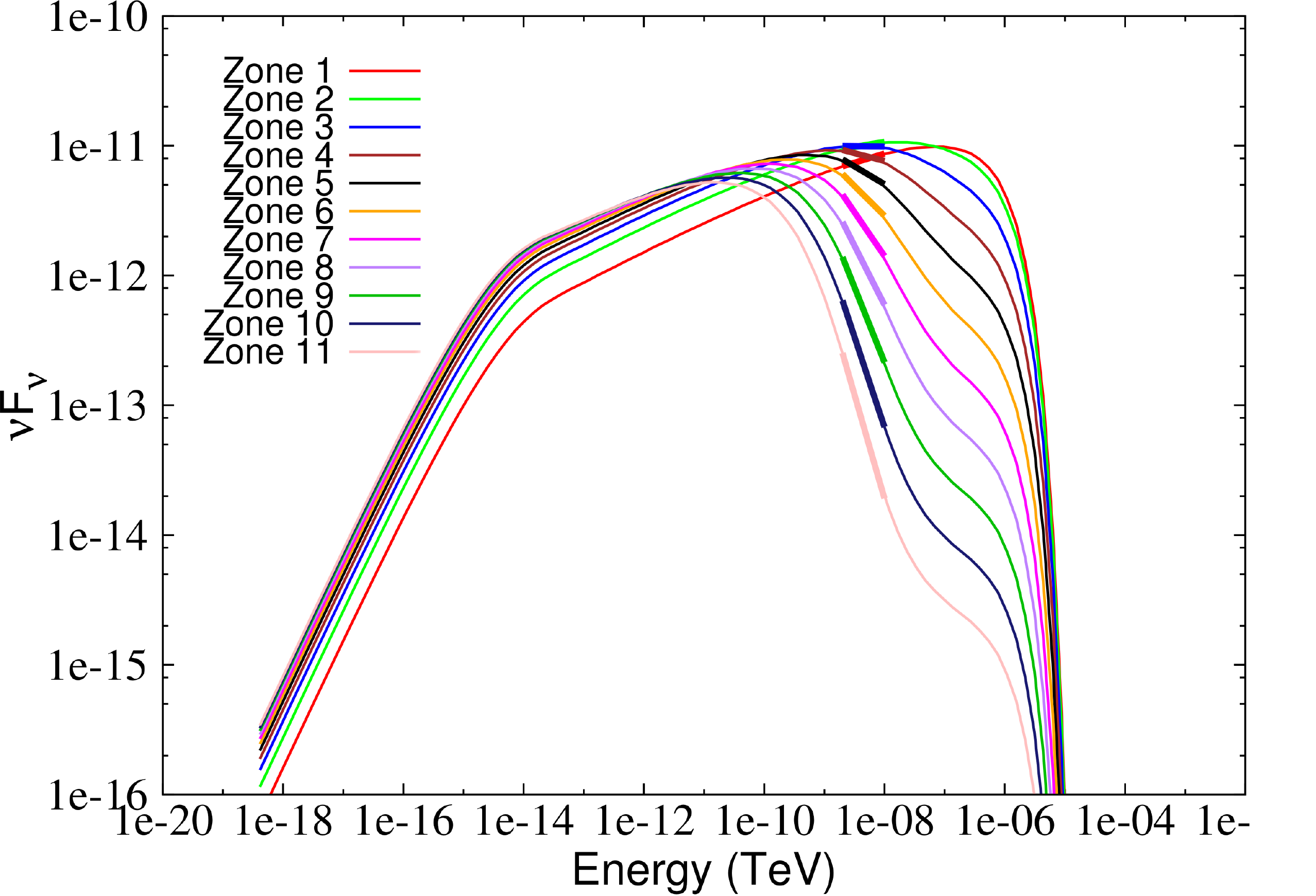}
  \caption{SR spectrum for PWN G21.5$-$0.9 for the first 11 LOS integration zones, with the spectral index in the X-ray energy band ($2 - 10$~keV) indicated by the thick lines for the by-eye best fit. 
    \label{Fig:G21_linesL}}
\end{figure}

\begin{figure}
  \centering
  \includegraphics[width=\columnwidth]{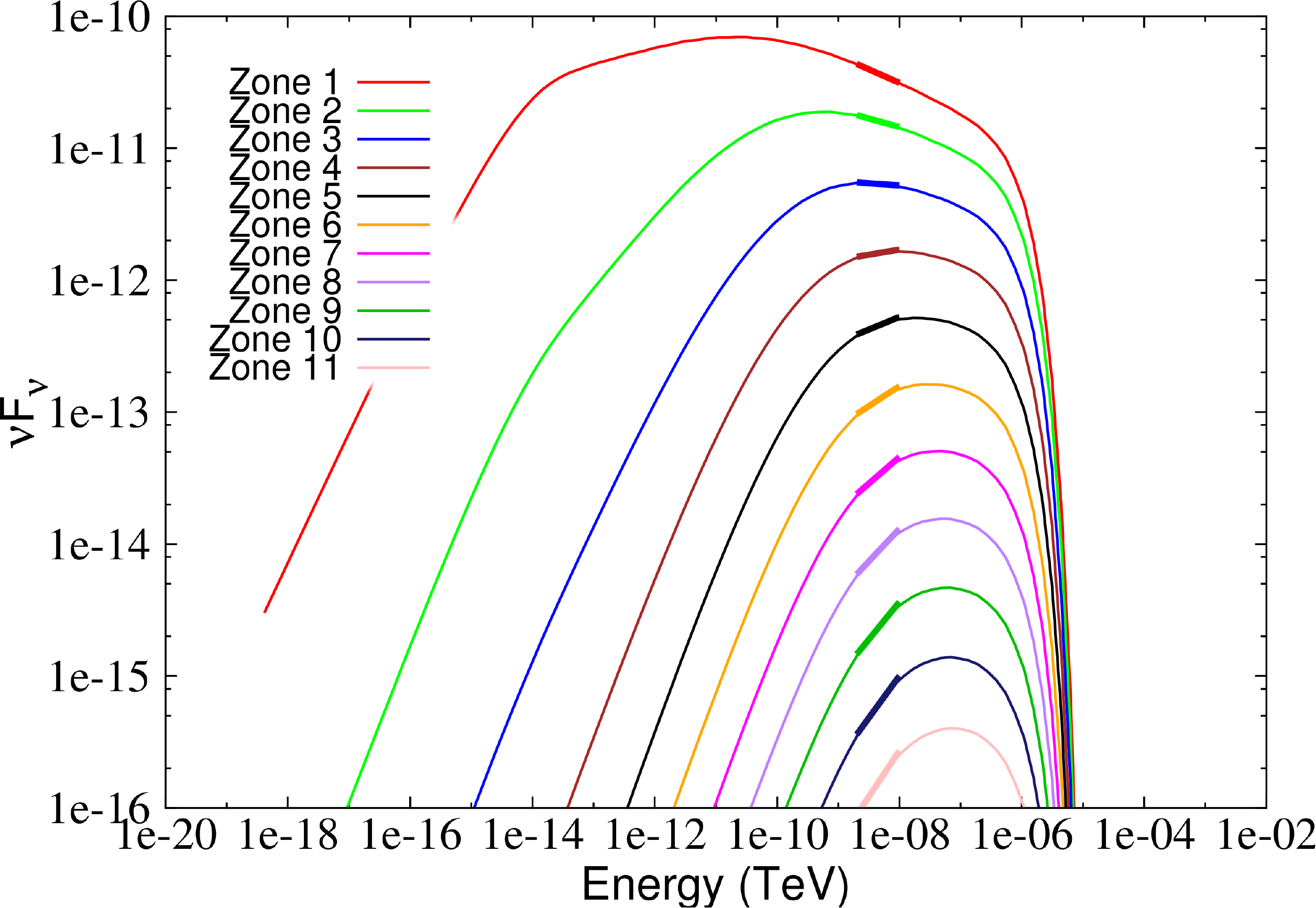}
  \caption{SR spectrum for PWN G21.5$-$0.9 for the first 11 LOS integration zones with the spectral index in the X-ray energy band ($2 - 10$~keV) indicated by the thick lines for the SFN best fit. 
    \label{Fig:G21_linesR}}
\end{figure}

In contrast to PWN 3C~58, PWN G21.5$-$0.9 did not give similar best fits for the three different fitting methods. In this source a form of degeneracy in the model becomes evident. From Figure~\ref{Fig:G21_SED} we see that all the methods give reasonable fits to the overall SED with the $\chi^2$ and the SFN methods not being able to fit the infrared data as well as the by-eye method. This can be attributed to the large errors on the infrared data that cause the $\chi^2$ and the SFN methods to assign a very small weight to these points. The by-eye method yielded a good fit to the SED as well as the X-ray spectral index profile, but was unable to find a simultaneous good fit to the SB profile. This resulted in the very large $\chi^2$ value for the by-eye method, as well as the very small $\chi^2_\Phi$ value in Table~\ref{tbl:G21} ($\chi^2_\Phi \ll 1$ is equivalent to $\chi^2/\nu \gg1$, indicating an extremely bad fit). This failure to find a good SB profile is due to the fact that the by-eye method favours a bulk speed normalisation that is a few orders of magnitude larger than that preferred by the other methods. This leads to the SB profile not decreasing as rapidly as the data indicate. The X-ray index profile, however, is particularly well fitted using this method. The degeneracy of the model becomes clear when one considers Figure~\ref{Fig:G21_SB} that shows that both the $\chi^2$ and SFN methods give similar good results in fitting the SB profile, but both being unable to fit the X-ray steepening as seen in the data (Figure~\ref{Fig:G21_index}). This is, however, a better overall fit with the combined $\chi^2$ value being much smaller than that of the by-eye method, as well as the $\chi^2_\Phi$ value being closer to $1.0$. The X-ray steepening has historically been attributed to a cooling effect\footnote{The SR energy loss rate is proportional to $E_{\rm e}^2$. Thus, higher-energy particles lose energy more rapidly as they move farther away from the central part of the PWN. This results in spectral steepening with distance, causing the X-ray photon index to increase with radius. If SR losses dominate, this results in spectral steepening.}. The best fits for $\chi^2$ and SFN, however, indicate the opposite trend, which might be interpreted as some sort of acceleration that is occurring in the system. This is, however, not the case here. We plot the SR spectrum for the first 11 radial bins and for two different parameter sets as found using the by-eye (Figure~\ref{Fig:G21_linesL}) and SFN (Figure~\ref{Fig:G21_linesR}) methods. The spectral index plotted in Figure~\ref{Fig:G21_index} is derived from the slopes of the thicker lines in the $2- 10$~keV energy range as shown in Figure~\ref{Fig:G21_linesL} and Figure~\ref{Fig:G21_linesR}. The expected cooling effect can be clearly seen in Figure~\ref{Fig:G21_linesL}, but in Figure~\ref{Fig:G21_linesR} the entire spectrum decreases rapidly as one moves away from the centre. This is due to the fact that in this case, the bulk flow of the particles are very slow and thus confines the particles to the inner parts of the PWN. The particles radiate most of their energy there before being able to move to the outer reaches of the PWN, leading to a rapid decrease in particle density with radius. This explains the hardening of the X-ray photon index with radius. In summary, we note that for this source we can either fit the SED and SB profile or the SED and X-ray spectral index profile, but not all three observables concurrently. Our inferred parameters are not too different from those of \citet{Porth2016} and \citet{Lu2017_3C}.

\begin{table}
\resizebox{\columnwidth}{!}{%
\begin{tabular}{lrrrr}
\hline
\textit{Fixed parameters}					& 	 		& 		&		&   \\ \hline 
Period ($P$) (s)					& & & 0.06186     		   		 		&   \\ 
Time derivative of period ($\dot P$) (s s$^{-1}$) 	& & & 2.0$\times10^{-13}$     		 		&   \\ 
Spin-down luminosity ($L_{\rm{age}}$) (erg/s)   	& & & 3.4$\times10^{37}$	 		 		&   \\ 
Braking index ($n$)               	        	& & & 3          		 		 		&   \\ 
Distance to the source (kpc)          			& & & 4.7          		 		 		&   \\ 
Index of the injected spectrum ($\alpha_1$) 		& & & 1.0         		 		 		&   \\ 
Index of the injected spectrum ($\alpha_2$) 		& & & 2.5	          	 		 		&   \\ 
Break energy ($\gamma_{\rm{b}}$) 			& & & 1.1$\times10^{5}$	 		 		&   \\ 
Conversion efficiency ($\eta$)               		& & & 0.01         		  		 		&   \\ 
Magnetic field time dependence ($\beta_{\rm B}$)		& & & $-1.0$         		  		 		&   \\
Soft-photon components:	& & $T$ (K) & $u$ (eV/cm$^3$) &       		   		 		   \\
\quad CMB		& & 2.76 & 0.23       		   		 		&   \\
\quad Infrared		& & 35.0 & 3.4      		   		 		&   \\
\quad Optical & & 3500.0 &  5.0     		   		 		&   \\ \hline
							& By-eye 		& $\chi^2$ 	&	$\chi^2_{\Phi}$	&   \\ \hline \hline

\textit{Fitted parameters}					& 	 		& 		&		&   \\ \hline 
Radial parameter of the magnetic field       			& 0.0          		& 0.0		&0.0		&   \\ 
Present-day magnetic field ($\mu$G)			& 72.7        		& 104	&74.4		&   \\ 
Bulk flow normalisation ($ 10^{10}$ cm s$^{-1}$)	& 0.81        		& 7.0$\times10^{-6}$		& 1.3$\times10^{-5}$		&   \\ 
Age (kyr)                			& 1.121      		& 0.980	&1.589	&   \\ 
Diffusion coefficient normalisation ($\kappa_{\rm{0}}$) 				& 4.3         		& 14.0		&74.0		&   \\ 
$\chi^2/\rm{\nu}$ 			& 117212/38		& 1265/38 & 1306/38		&   \\
& (3085)		&  (33.2)&  (34.4)		&   \\
$\chi^2_\Phi$			& $-64.2$         		& 0.59		& 0.60		&   \\ \hline
\end{tabular}
}
\caption{Best-fit parameters for PWN G21.5$-$0.9.\label{tbl:G21}}
\end{table}

\subsection{PWN G0.9+0.1}\label{subsec:G0.9}
PWN G0.9+0.1 is a well-known composite SNR \citep{HelfandB1987}, 
as indicated by its characteristic radio morphology: it exhibits a flat-spectrum radio core ($\sim2'$ across) corresponding to the PWN, and also clear, steeper shell components ($\sim 8'$ diameter shell). 
This bright, extended source near the Galactic Centre has since become a well-known SNR, with an estimated age of a few thousand years \citep{G0.9+0.1_HESS} with a lower bound of $1100$ yr \citep{Dubner2008} and a typical distance of 8.5 kpc. During X-ray observations of the Galactic Centre, \cite{Sidoli2004} serendipitously observed SNR G0.9+0.1 using the \textit{XMM}-Newton telescope. \cite{Sidoli2004} fit an absorbed power-law spectrum that yielded a photon index of $\Gamma_{\rm X} \sim 1.9$ and an energy flux of $F = 4.8 \times 10^{-12}$ erg cm$^{-2}$ s$^{-1}$ in the 2$-$10 keV energy band. This translates to a luminosity of $L_{\rm X} \sim 5 \times 10^{34}$ erg s$^{-1}$ for a distance of 10 kpc. \cite{G0.9+0.1_HESS} obtained a a power-law fit to the observed $\gamma$-ray spectrum with a photon index of $2.29 \pm 0.14_{\rm{stat}}$ and an integral photon flux of $(5.5 \pm 0.8_{\rm{stat}})\times 10^{-12}$~cm$^{-2}$~s$^{-1}$ above 200~GeV. Subsequently, the radio pulsar PSR~J1747$-$2809 was discovered in PWN~G0.9+0.1 with a period $P = 52.2$~ms and $\dot{P} = 1.56\times10^{-13} \rm s~\rm s^{-1}$ \citep{G0.9+0.1data}.

\begin{figure}
  \centering
  \includegraphics[width=\columnwidth]{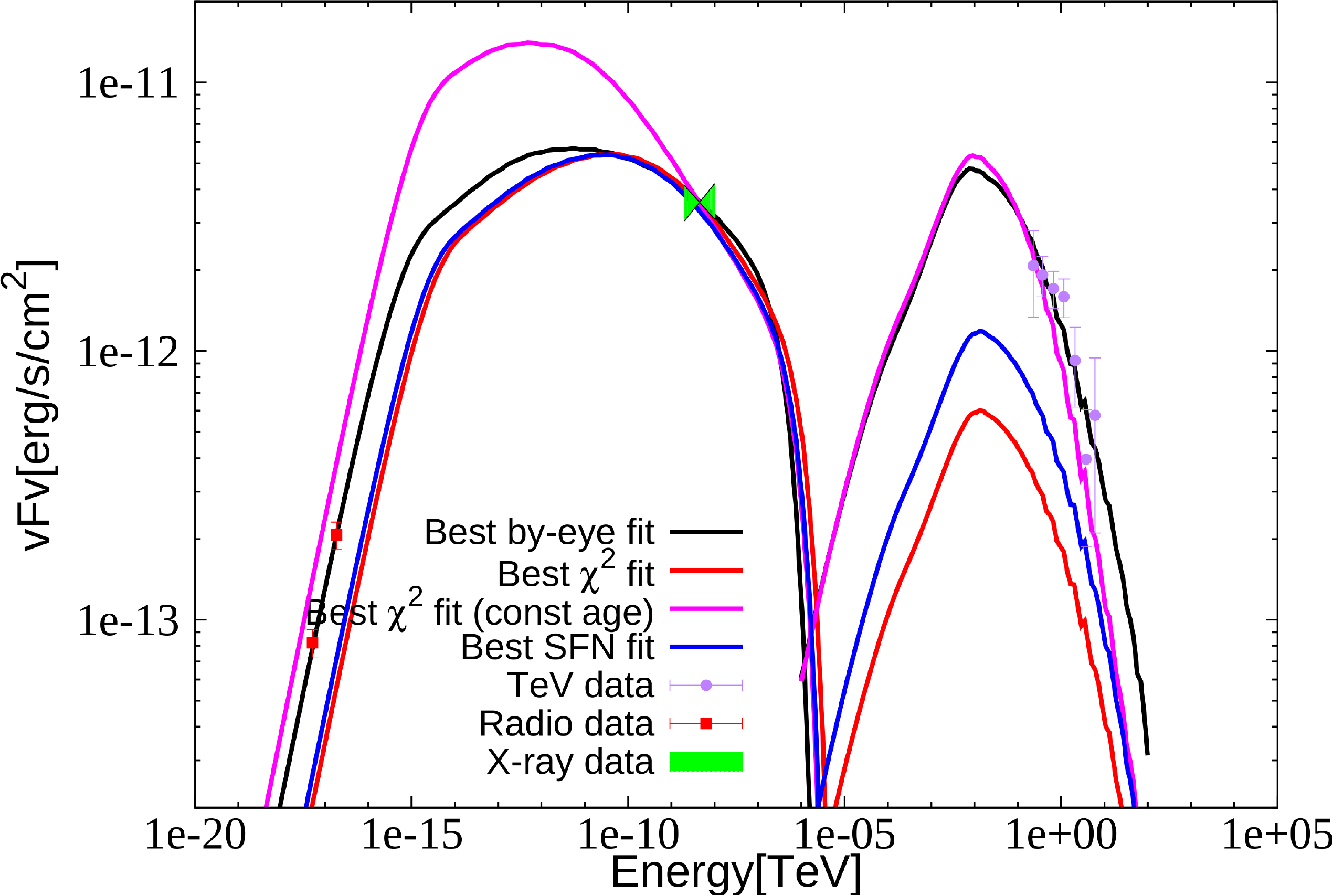}
  \caption{Broadband SED for PWN G0.9+0.1, with radio data from \citet{Dubner2008}, X-ray data from \citet{Porquet2003} and the TeV data from H.E.S.S.~\citep{G0.9+0.1_HESS}.\label{Fig:G09_SED}}
\end{figure}

\begin{figure}
  \centering
  \includegraphics[width=\columnwidth]{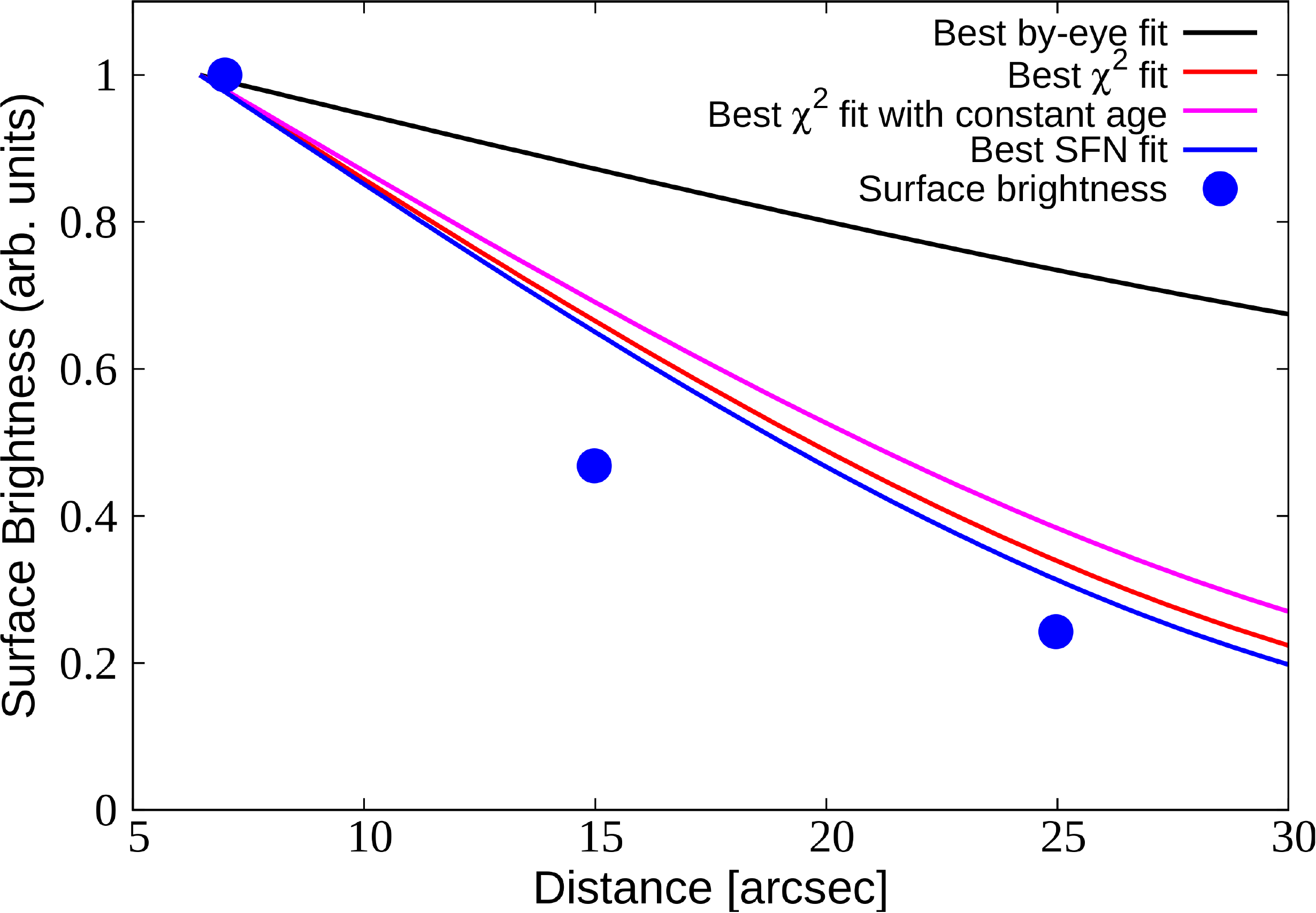}
  \caption{SB profile for PWN G0.9+0.1 with data points from \citet{2012XMMG09} and the lines indicating the model best fits.
    \label{Fig:G09_SB}}
\end{figure}

\begin{figure}
  \centering
  \includegraphics[width=\columnwidth]{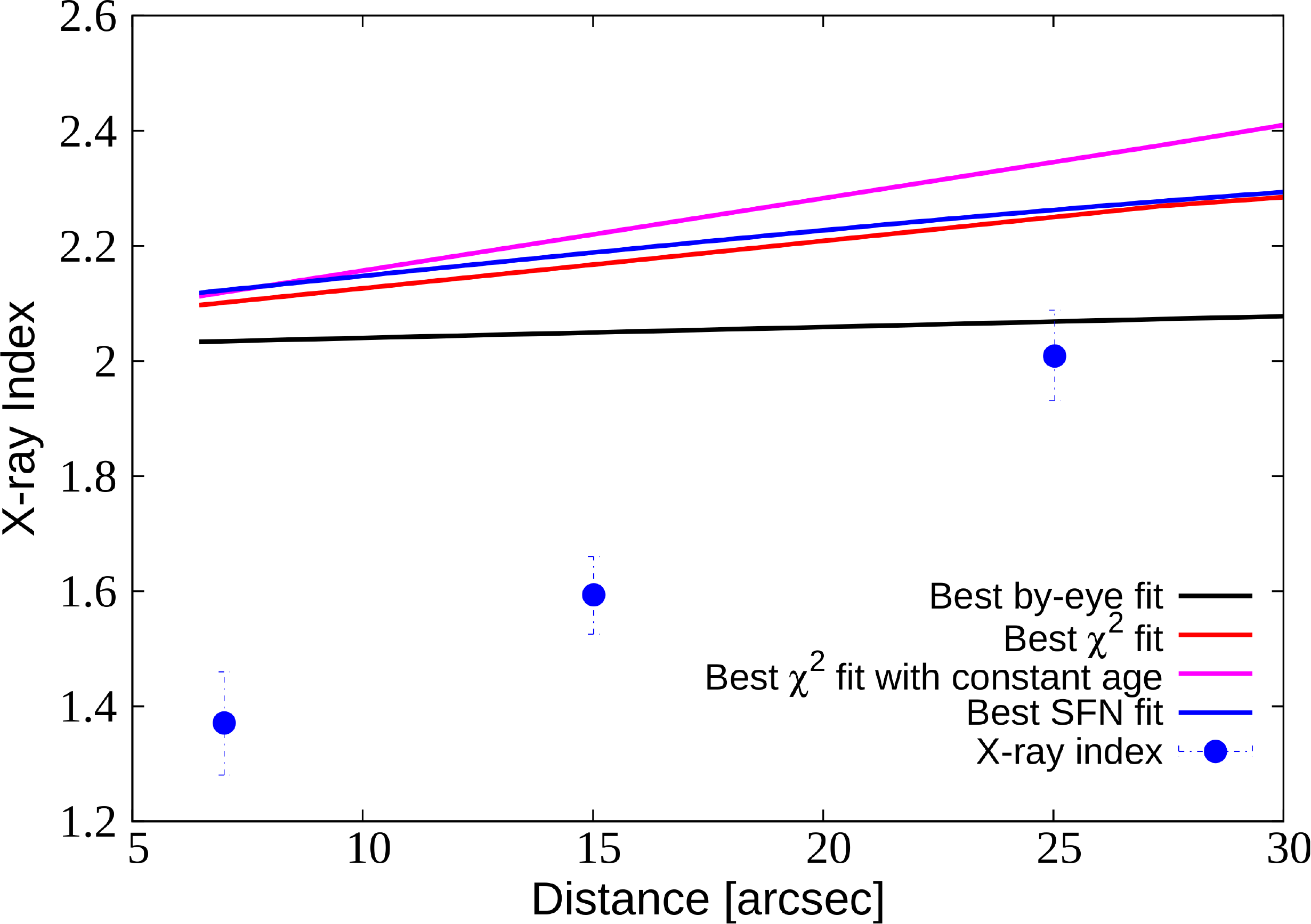}
  \caption{X-ray photon index profile for PWN G0.9+0.1 with data points from \citet{2012XMMG09} and the lines indicating the model best fits.
    \label{Fig:G09_index}}
\end{figure}

Figures~\ref{Fig:G09_SED} to~\ref{Fig:G09_index} show the results for PWN G0.9+0.1. The by-eye method was able to find a good fit to the SED and a reasonable fit to the X-ray index profile, but was unable to reproduce the SB profile, since (as in the previous case) the preferred bulk flow of the particles is large (driven by the X-ray photon index profile), resulting in a larger source. This led to rather poor goodness-of-fit values according to the $\chi^2$ and $\chi^2_\Phi$ test statistics. The $\chi^2$ and SFN methods were able to find reasonable fits to the SED and a comparatively better fit to the SB profile. Their goodness-of-fit values are relatively close, with each method preferring slightly different parameters. However, these fits preferred a younger age for the PWN, resulting in fewer particles in the system and thus a worse fit to the high-energy tail of the IC component of the SED. From the model of \citet{Torres2014}, the preferred age of the system is a couple of thousand years. We thus decided to fix the age for to 3~078 yr, which is the age preferred by the by-eye method and this agrees better with the pulsar age of 5~600 yr. We then redid the $\chi^2$ test, leaving the present-day magnetic field, the bulk flow of particles and the diffusion normalisation as free parameters. The result is indicated by the magenta line in Figures~\ref{Fig:G09_SED} to \ref{Fig:G09_index} and the best-fit values are shown in brackets in Table~\ref{tbl:G09}. In this case, we found a good fit to the SED as well as to the SB profile, and the trend in the X-ray index was more accurately represented, characterised by a reduced $\chi^2$ value of $19.6$ and a $\chi^2_\Phi$ value of $0.73$. In summary, the code was not able to find a simultaneous good fit to all three data subsets. 
PWN G0.9+0.1 has been modelled by \citet{VdeJager2007} and \citet{Torres2014}. \citet{VdeJager2007} found similar spectral results with the exception that their modelled age was 6.5~kyr, which is twice the age predicted by \citet{Torres2014} and by our by-eye method. The predicted present-day magnetic field vary slightly between the three modelling attempts with our predictions being the largest and \citet{VdeJager2007} being the smallest, but all the predictions are in the order of tens of $\mu \rm G$.

\begin{table}
\resizebox{\columnwidth}{!}{%
\begin{tabular}{lrrrr}
\hline
\textit{Fixed parameters}					& 	 		& 		&		&   \\ \hline 
Period ($P$) (s)					& & & 0.0522     		   		 		&   \\ 
Time derivative of period ($\dot P$) (s s$^{-1}$) 	& & & 1.56$\times10^{-13}$     		 		&   \\ 
Spin-down luminosity ($L_{\rm{age}}$) (erg/s)   	& & & 4.3$\times10^{37}$	 		 		&   \\ 
Braking index ($n$)               	        	& & & 3          		 		 		&   \\ 
Distance to the source (kpc)          			& & & 8.5          		 		 		&   \\ 
Index of the injected spectrum ($\alpha_1$) 		& & & 1.4         		 		 		&   \\ 
Index of the injected spectrum ($\alpha_2$) 		& & & 2.7	          	 		 		&   \\ 
Break energy ($\gamma_{\rm{b}}$) 			& & & 1.0$\times10^{5}$	 		 		&   \\ 
Conversion efficiency ($\eta$)               		& & & 0.01         		  		 		&   \\ 
Magnetic field time dependence ($\beta_{\rm B}$)		& & & $-1.0$         		  		 		&   \\ 
Soft-photon components:	& & $T$ (K) & $u$ (eV/cm$^3$) &       		   		 		   \\ 
\quad CMB		& & 2.76 & 0.23       		   		 		&   \\
\quad Infrared		& & 30.0 & 2.5      		   		 		&   \\
\quad Optical & & 3000.0 &  25.0     		   		 		&   \\ \hline
							& By-eye 		& $\chi^2$ 	&	$\chi^2_{\Phi}$	&   \\ \hline \hline
\textit{Fitted parameters}				& 	 		& 		&		&   \\ \hline 
Radial parameter of the magnetic field       	& 0.0          		& 0.0		&0.0		&   \\ 
Present-day magnetic field ($\mu$G)				& 15.0        		& 37.7 (25.1)		&26.8		&   \\ 
Bulk flow normalisation ($10^{10}$ cm s$^{-1}$)	& 0.27        		& 9.5$\times10^{-2}$ (5.2$\times10^{-2}$) &4.5$\times10^{-2}$		&   \\ 
Age (kyr)                			& 3.078      		& 0.611 (3.078)	&1.085	&   \\ 
Diffusion coefficient normalisation ($\kappa_{\rm{0}}$) 				& 2.2         		& 0.92 (0.11)		&0.35		&   \\
$\chi^2/\rm{\nu}$ 			& 900/32 & 570/32 & 762/32  &   \\
	&  (28.1)		& (17.8)	&  (23.8)		&   \\
$\chi^2_\Phi$			& 0.24         		& 0.75		& 0.76			&   \\ \hline
\end{tabular}
}
\caption{Best-fit parameters for PWN G0.9+0.1.\label{tbl:G09}}
\end{table}

\subsection{Characterising the Non-colocation of the Various Best-fit Solutions}
We find that our different search methods yield different, i.e., non-unique best-fit parameters. This is because: (1) the statistical metric (or by-eye intuition) used to assess the goodness of fit in each case, differs;
(2) the methods use different ways to combine the different data sets in order to find a compromise solution that concurrently fit all data.
 
Ideally, all methods would find the same answer, and no compromise would have been needed when fitting multiple data sets. Also, the value of $\chi^2_\Phi$ would have been unity for both the single sets and the average value. However, since the optimal solutions differ between methods, it would  be good to quantify this discrepancy. Using parameter errors may indicate whether the contours found by different methods overlap; however, these primarily reflect observational errors and may not be the best characterisation of this kind of model degeneracy. 
 
A more relevant proxy of ``non-colocation'' of the respective best-fit parameters chosen by each method may be to quantify the amount of compromise in goodness of fit that occurs when fitting a single data set vs.\ concurrently fitting all three sets. 

\begin{table}
\resizebox{\columnwidth}{!}{%
\begin{tabular}{lrrrrrrrrr}
& $B_{\rm{age}}$    & Age & $V_0$ &     $\kappa_{\rm{0}}$      &     $\chi^2_\Phi$ &     $\chi^2_\Phi$ &     $\chi^2_\Phi$  &     $\chi^2_\Phi$ &\\
& ($\mu$G) & (yr) & ($10^{10}$ cm s$^{-1}$) & & (SED) & ($\Gamma_{\rm X}$) & (SB) & (Avg.)\\
\hline
\textit{PWN 3C 58} & & & & & & & &\\
\hline
Global & 39.1    &   1588.9  &   1.02 &  74.0 & 0.99    &   0.92    &   0.92    &   0.94    &\\
SED & 50.4    &    743.0  &   1.58 &   4.2 & \textbf{0.99}    &  0.91    &   0.61    &   0.84    &\\
$\Gamma_{\rm X}$ & 81.9    &   2134.0  &   0.71 &  13.9 & 0.53    &   \textbf{0.94}    &   0.45    &   0.64    &\\
SB & 43.6    &   2114.0  &   0.22 & 104.7 & 0.1     &   0.88    &   \textbf{0.93}    &   0.64    &\\
By-eye & 65.7    &   1121.0  &   1.8 &  81 & 0.99    &   0.94    &   0.72    &   0.88    &\\
\hline

\end{tabular}
}
\caption{Best-fit parameters found when fitting all data sets concurrently (first row for each source) vs.\ only fitting one data set at a time (subsequent rows) for PWN 3C~58, with the relevant $\chi^2_\Phi$ values indicated in boldface and the implied values for the other sets also shown. The average $\chi^2_\Phi$ value is given in the final column.  
\label{tbl:co-loc}}
\end{table}
Table~\ref{tbl:co-loc} indicates the $\chi^2_\Phi$ values for four different scenarios, for each modelled source. The first scenario (labelled ``Global'') is for fitting all three data sets concurrently, and is shown in the first row. The next three scenarios are when a single data set is fit without regarding any other set; the relevant $\chi^2_\Phi$ value is shown in boldface for the individual data sets and is labelled ``$\chi^2_\Phi$ (SED)'', ``$\chi^2_\Phi$ ($\Gamma_{\rm X}$)'' and ``$\chi^2_\Phi$ (SB)'', respectively. Also indicated are the implied values of  $\chi^2_\Phi$ for the other data sets, since we do have a model prediction for these data even if they were not explicitly fitted. Lastly, the average value for $\chi^2_\Phi$ is shown in the final column.

From the table, one can see that the global fit yields the best average $\chi^2_\Phi$ value, and all other fits involving only a single data set have lower average values, indicating an optimal compromise in the first case. Second, when only a single data set is used for a particular fit, the single-set value of $\chi^2_\Phi$ may be high, but the corresponding fits for the other sets are not good. 
So, in the first source for example, the non-colocation of the respective single-set best fit parameters leads to a compromise concurrent fit, with $\chi^2_\Phi = 0.94$ for this global fit. 
We thus move away from the optimal single-set fits with values of 0.99, 0.94 and 0.93 and obtain a global compromise solution with values of 0.99, 0.92 and 0.92. 
This represents a slight loss in goodness of fit of  $\Delta\chi^2_\Phi = 0.01$ with respect to the ideal single-set fits, for the last two data sets. In the ideal case, the global fit would thus have had an average value of $\chi^2_\Phi=0.95$ (average of bold table entries) instead of the 0.94 we now obtain, indicating only a slight compromise for the global fit, but still obtaining the highest average value of $\chi^2_\Phi$ compared to the other rows in the table.

The $\chi^2_\Phi$ values for the by-eye fit are usually not good, and underscores that a by-eye method uses a different intuitive metric that may not be regarded as statistically optimal, but may yield model solutions that can be used as a basis for comparison with the fits yielded by the statistical methods. In addition, we found that when we perform the exercise described above for the other two sources, the single-set fits have problems converging, and the values for $\chi^2_\Phi$ are quite bad for the other bands. This indicates that we need the compromise fit to have any hope of a reasonable fit to all data sets. Future model refinement should help to minimise the effect of non-colocation of single-band best-fit parameters.

\section{Conclusion}\label{sec:concl}
In this paper we presented results from a spherically-symmetric, spatially-dependent particle transport and emission code for young PWNe. We note that convection and SR losses dominate the physics, indicating the importance of constraining the bulk motion and magnetic field profiles. We were able to predict the SEDs, SB profiles and X-ray photon indices vs.\ radius for three PWNe. Our model found reasonable concurrent fits to these observables, but each source posed its own challenges. We found that a spatially constant magnetic field was preferred (explaining the success of the 0D models that assume a spatially constant magnetic field; e.g., \citealt{Gelfand2009}), leading to a velocity profile that scales as $V(r)\sim 1/r$. Our code was also able to predict spectral steepening with increasing radius due to cooling via SR losses. 

We followed various approaches to obtain best fits to the available data, finding that not all methods give similar best fits. 
They do, however, shed light on the problems that arise when fitting heterogeneous data, indicating that the constraining power of the data on the models under scrutiny is also subject to the fitting methods one chooses to use. A by-eye fitting method proved useful, but is limited by the size of the parameter space one needs to explore. The $\chi^2$ test statistic yielded reasonable results, but in this framework data with small relative errors dominate other data sets, practically eliminating the effect of the latter when minimising the $\chi^2$ value. The SFN test statistic proved advantageous when dealing with heterogeneous data sets. However, we could not clearly prefer one of these methods over the others.
In particular, non-location of best-fit parameters preferred by each separate data set lays bare some model degeneracies, since in the ideal case parameter fits on the various individual data sets should yield consistent results when performing concurrent fitting of all sets involved.
This may point not only to the power of each search method, but perhaps to a revision that is needed in the model, since any fitting method would ``fail'' to find adequate solutions if such solutions do not exist in the model's solution space in the first place. Since the parameter space for these types of models are complicated and large, it would be worthwhile to explore it in more detail as this would help gauge the uniqueness of best-fit solutions as well as degeneracies that exist between some parameters. This can be done by using, e.g., an MCMC ensemble sampler for a fine spatial and energy resolution to give a better understanding of the parameter space properties and could help to estimate errors on best-fit parameters. 

Our best-fit parameters (notably magnetic fields, bulk flow normalisations, and diffusion coefficients) resemble those found by \citet{Porth2016} and \citet{Lu2017_3C}, although relatively wide ranges are allowed for several of these. On the one hand, this indicates some consistency in the independent approaches, but on the other hand also some degeneracy in the sense that the currently available data do not yet have the discriminatory power to distinguish between models that make quite different assumptions about the PWN environment. We developed our code to have as few free parameters as possible, thus assuming very basic profiles for, e.g., the magnetic field and the bulk particle motion. Model degeneracy will have to be broken by future observations, including polarisation properties that may better constrain the magnetic field topology.  

As alluded to above, a second type of degeneracy was uncovered while fitting the spectral and spatial data of G21.5$-$0.9 and G0.9+0.1. We could fit two out of three data subsets, but not all three subsets in two of the three PWNe we considered. This probably points to some physics that we are missing in the current model. One avenue would be to refine the parametrisation of our model magnetic field and bulk flows, or to connect them in a different way so that we can allow more freedom in the model to fit both the SB  and the X-ray index profiles, in addition to the SED. We could also reassess our choice of free parameters and attempt to more closely incorporate MHD results in our model calculations. Other ideas include the expansion of our code to more spatial dimensions, or invoking a spatially-dependent particle injection spectrum (possibly including co-latitudinal dependence to reflect recent results in pulsar wind simulations, e.g. \citealt{Tchek16}). In addition, it could be fruitful to more rigorously include the dynamical evolution of the PWN-SNR system, so that we can also model older PWNe in future \citep{Martin2016}.

Continued improvement both in the PWN model and in fitting methods will help us make the best use of forthcoming morphological data, especially in the VHE waveband. Adding different types of data using one consistent framework would also increase the constraining power thereof.

\section*{Acknowledgements} We thank the referee whose insightful comments helped improve the paper significantly. We also thank Zorawar Wadiasingh and Matthew Baring for fruitful discussions.
This work is based on the research supported wholly / in part by the National Research Foundation (NRF) of South Africa (Grant Numbers 92860, 93278, and 99072). The Grantholder acknowledges that opinions, findings and conclusions or recommendations expressed in any publication generated by the NRF supported research is that of the author(s), and that the NRF accepts no liability whatsoever in this regard.

\label{Bibliography}
\bibliographystyle{mnras}  
\bibliography{Bibliography}

\end{document}